\documentclass[12pt,preprint]{aastex7}

\usepackage{graphicx}
\usepackage{amsmath}
\usepackage{amssymb}
\usepackage{lineno}

\definecolor{purple}{rgb}{1,0,1}

\shorttitle{Magnetar Pair Cascades}
\shortauthors{Harding et al.}
\graphicspath{{./}{figures/}}

\def\be{\begin{equation}}
\def\ee{\end{equation}}

\def\lambar{\lambda\llap {--}}

\def\tt{\mbox{\boldmath $\theta $}}

\def\lambar{\lambda\llap {--}}

\def\lsim{\lower 2pt \hbox{$\, \buildrel {\scriptstyle <}\over
         {\scriptstyle \sim}\,$}}
\newcommand\gsim{\buildrel > \over \sim}

\begin{document}

\title{\uppercase{Pair Cascades in Magnetar Magnetospheres}}

\author[0000-0001-6119-859X]{Alice K. Harding}
\affiliation{Theoretical Division, Los Alamos National Laboratory, Los Alamos, NM 87545}
\email{ahardingx@yahoo.com}
\author[0000-0002-9249-0515]{Zorawar Wadiasingh}
\affiliation{Department of Astronomy, University of Maryland, College Park, Maryland 20742, USA}
\affiliation{Astrophysics Science Division, NASA Goddard Space Flight Center,Greenbelt, MD 20771, USA}
\affiliation{Center for Research and Exploration in Space Science and Technology, NASA/GSFC, Greenbelt, Maryland 20771, USA}
\email{}

\author[0000-0003-4433-1365]{Matthew G. Baring}
\affiliation{Department of Physics and Astronomy—MS 108, Rice University, 6100 Main St., Houston, TX 77251-1892, USA}
\email{}


\begin{abstract}
Resonant inverse Compton scattering (RICS) of soft thermal photons by relativistic particles on closed magnetic field loops has been proposed to explain the hard emission observed up to, and beyond, 200 keV from magnetars.  If particles injected at the base of the loops have Lorentz factors $\gsim 10^2$, the RICS spectra will be attenuated by both one-photon pair production and photon splitting in the ultra-strong magnetar fields, producing additional spectral components from pair synchrotron radiation and split photons that produce further generations of pairs and split photons.  We investigate such cascades initiated by the primary injected electrons through a Monte Carlo simulation and study the cascade spectra and pair distributions.  For most observer angles, the pair synchrotron and split photon spectra dominate the RICS primary spectra and produce complex polarization signals.  In particular, the synchrotron spectra are highly polarized with degree 40\% - 80\%, are softer than the RICS spectra and may account for the high polarization of some magnetar spectra observed by IXPE above 3 keV.  
\end{abstract}

\keywords{}


\defcitealias{Wadiasingh2018}{W2018}
\section{Introduction} \label{sec:intro}

Magnetars, bursting neutron stars that appear to be magnetically powered, possess the strongest known magnetic fields of any astrophysical source.  Their fields inferred from dipole spin-down are mostly in the range $10^{14} - 10^{15}$ G but the actual surface field strengths may be higher and the interior field strengths higher still.  Such high fields are well above the quantum critical field strength, $B_{\rm cr} = 4.413 \times 10^{13}$ G, enabling exotic QED processes such as photon splitting, and requiring QED treatment of all physical processes.  There are currently some 30 known magnetars, which have periods of $P~=~1~-~12$~ s~\citep{2014ApJS..212....6O}, although there have been recent discoveries of potential ultralong-period magnetars with $> 1$ min up to several hours \citep{2022NatAs...6..828C,Hurley2022,2023MNRAS.520.1872B,2023Natur.619..487H,2024NatAs...8.1159C,2024arXiv241116606W,2025arXiv250109133L,2025arXiv250110528M}, including the hard X-ray emitter 1E 161348--5055 \citep[e.g.,][]{2006Sci...313..814D,2016MNRAS.463.2394D,2016ApJ...828L..13R,2018MNRAS.478..741B} in RCW 103.  The relatively long periods give dipole spin-down powers that are orders-of-magnitude less than their persistent emission luminosities of $L_{\rm x} \sim 10^{35} - 10^{36}\, {\rm erg\,s^{-1}}$, most of which is seen at X-ray wavelengths.   The observed persistent spectra consist of a thermal component with temperatures $\sim 0.5$ keV, a soft power law with photon index $1.5 - 4$ seen up to $10-20$ keV, and a hard power law with photon index $0.3 - 1.7$ \citep{Enoto2010} seen to at least 100 - 200 keV.

The energy source for magnetars is believed to be the decay of the magnetic field which causes re-arrangement of near surface fields, leading to crustal motions \citep{Pons2009}.  Since the external magnetic field is anchored to the (mobile) ionic lattice in the crust, such motions can cause local twists of the magnetic field that induce magnetospheric currents \citep{Thompson2002}.  The decay of the twists in the global dipole field produces electric fields that can accelerate particles along dipole field loops \citep{Belo2007,Belo2013a}.  The magnetic field twists are sustained by currents, and the particle acceleration must self-consistently supply that current through creation of electron-positron pairs \citep{Chen2017}.  The accelerated particles also radiate some of their energy, potentially producing the observed, persistent hard X-ray emission.  \citet{Belo2013b} modeled the emission and production of pair plasma on twisted, closed field loops.  If the Lorentz factors of the particles reach $10 - 10^3$, the primary source of radiation will be resonant inverse Compton scattering (RICS) of the thermal photons from the neutron star \citep{BaringHarding2007} within a few stellar radii above the surface.  The  RICS photons will undergo both one-photon pair production and photon splitting in the strong magnetic field, depending on their polarization \citep{BaringHarding2001}, initiating a pair cascade of multiple generations.  

The pair cascade calculations of \citet{Belo2013b}, while the first such study for magnetars, made a number of assumptions.  Classical Compton scattering cross sections were used.   All pairs were assumed to be produced at one-photon pair threshold  in the ground Landau state, which is accurate in the very high, near-surface magnetic fields but not accurate at several stellar radii where the field drops below $B' \simeq B/B_{\rm cr} \lsim 0.1$ \citep{BaringHarding2001}.  Such an assumption prevents any synchrotron radiation from the pairs.  Finally, the thermal target photons were assumed to be mono-energetic, which significantly impacts where and what particle energies and photon angles the resonance is accessed \citep{Baring2011,Cooper2024}.
We will present a pair cascade simulation on closed dipole field loops in a magnetar magnetosphere that makes none of these assumptions.  We will show that allowing the production of pairs in excited Landau states enables robust cascades with multiple pair generations, producing significant pair synchrotron radiation components that overwhelm the RICS radiation primarily at around a few MeV and even lower energies.

Our study is based on previous models of RICS radiation from particles on single closed magnetic field loops, notably that of \citet{Wadiasingh2018}, hereafter \citetalias{Wadiasingh2018}) who used QED Compton scattering cross sections appropriate for the high magnetar magnetic fields.  They computed RICS spectra for particles that experience no radiative energy loss and did not include spectral attenuation by photon splitting or pair production.  They found that the uncooled RICS spectra from single field lines are very hard compared to those of observed magnetar hard components and that their shape and cutoff are strongly dependent on observer viewing angle to the magnetic field axis.  Wadiasingh et al. (2025) expand on this work by computing RICS spectra integrated over volumes of closed field loops and included attenuation by pair production and photon splitting.  They find that the integrated spectra (still from uncooled electrons) are much softer than the spectra for single loops, and agree much better with the observed hard-component spectra seen above 10 keV \citep[e.g.,][]{Kuiper-2004-ApJ,denHartog-2008-AandA,Enoto2010,2017ApJS..231....8E}.  They did, however, not take into account the additional radiation produced by the pairs and split photons participating in cascades.

In this paper, we present a Monte-Carlo simulation of a pair cascade that is initiated by RICS emission from a single closed magnetic field loop.  We include RICS radiation losses of the particles that is significant at magnetar field strengths.  A single-loop instead of a volume-integrated approach is chosen since it is easier to study the effects of the RICS losses and cascade development alone, and to compare the RICS loss cases with the uncooled cases presented in W2018.  We therefore will add the effects of cascades on multiple field loops in a future study.  In Section~\ref{sec:model}, we discuss the physics and physical processes that are included in the simulation as well as a description of the Monte-Carlo (MC) code (MAGCAS).
Section~\ref{sec:results} presents the results for the RICS spectra in Section~\ref{sec:RICS_spectra} and cascade spectra in Section~\ref{sec:cas_spectra}.  The distribution of created pairs and their spectra are shown in Section~\ref{sec:pairs}.  Finally in Section~\ref{sec:disc} we discuss the implications of our results for magnetar spectral and polarization measurements.

\section{Pair Cascade Simulation} \label{sec:model}

Electromagnetic pair cascades have been studied for many years as the source of plasma for pulsar magnetospheres \citep{Daugherty1982}.  More recently, the pair cascades that occur near the polar caps are seen as necessary for supplying both the global magnetosphere currents \citep{2010MNRAS.408.2092T,Timokhin2013} and the production of coherent radio emission \citep{Philippov2020,2024ApJ...974L..32C,2024A&A...691A.137B,2025arXiv250317249B}.    In pulsar pair cascades, the main source of photons is curvature radiation and non-resonant inverse-Compton emission, and pairs are created by one-photon pair production in excited Landau states and produce abundant synchrotron photons that spawn many pair generations \citep{Timokhin2015}.  RICS can dominate losses very near the surfaces of pulsar polar caps, at Lorentz factors $\gamma\sim 10^2 - 10^3$ but the electric fields are high enough that particles can continue accelerating to higher energies of $\sim 10^7$, where curvature radiation dominates over RICS \citep{Harding1998} because of the strong dependence of its cooling rate $\dot{\gamma}_{\rm curv} \propto \gamma^4$ on the electron Lorentz factor.  In magnetar magnetospheres, the much higher magnetic fields and neutron star (NS) surface temperatures greatly increase RICS loss rates, preventing particles from reaching energies high enough for curvature radiation, except perhaps in cases of extreme twists with very high electric fields \citep{Cooper2024}.  Magnetar pair cascades will thus have a very different character \citep{BaringHarding2001}.  In the strongest fields near the NS surface, photon splitting will dominate over pair production at least for photons with electric vectors perpendicular to the plane of the local magnetic field and the photon momentum ($\perp$ mode, if splitting takes place only for the $\perp \rightarrow \parallel\parallel$ mode, more details are given in Section~\ref{sec:split}).  Photons with electric vectors parallel to that plane ($\parallel$ mode) will produce pairs at threshold, $2m_ec^2/\sin\theta_{\rm kB}$, where $\theta_{\rm kB}$ is the angle between the photon momentum and the local magnetic field, and in the ground Landau state such that synchrotron radiation does not occur.  For radiation by particles on closed field loops and depending on the surface magnetic field strength, the local magnetic field may decrease below $10^{13}$ G where photon splitting no longer dominates and photons of both polarization modes can produce pairs.  As the field decreases further, pairs are produced in higher Landau states, allowing more synchrotron photons and more pair cascade generations, bolstering the emission at sub-MeV energies.  

In the following sections, we describe the processes that are included in our MC simulation as well as the assumptions that are made.  The MC code is described in Section~\ref{sec:MC} as well as the geometry of the emission and cascade development.

\subsection{Resonant Inverse Compton Emission}  \label{sec:RICS}

We will assume that primary particles injected at the NS surface are accelerated by twisted loops of equatorial dipole field lines and scatter soft thermal photons.  As the particles move along closed field loops, the scattered photons are directed along the local magnetic field direction $\hat{\boldsymbol{B}}$ toward observers at an angle, $\theta_v = \arccos (\hat{z} \cdot \hat{\boldsymbol{B}})$, to a dipole axis aligned with the $z$ axis.  To compute the scattered photon spectrum for both polarization modes at each step along the particle trajectory, we use an adapted form of Eqn (17) of \citetalias{Wadiasingh2018}
\be \label{eqn:ngamma}
{dn_\gamma^{\perp,\parallel}(\varepsilon_f, \theta_v) \over dtd\varepsilon_f} = {c \over 2\pi \gamma^3 \beta \,\Delta S (\varepsilon_f)}\, \int ds {\hat\omega_i \over 1+\beta\mu_B}\,\left|{\partial \omega_i \over \partial \omega_f}\right|
{d\sigma^{\perp,\parallel} \over d\cos\theta_f} \int_{\varepsilon_-}^{\varepsilon_+}\,{d\varepsilon_s \over \varepsilon_s^2}\,n_\gamma(\varepsilon_s)\,f(\mu_i)
\ee
where $\gamma$ is the particle Lorentz factor, $\varepsilon_i$ and $\varepsilon_f$ are the incident and scattered photon energies in the observer frame (OF), $\omega_i$ and $\omega_f$ are the incident and scattered photon energies in the electron rest frame (ERF), $\mu_i$ is the cosine of the angle between the incident photon momentum and local magnetic field in the OF.  Hereafter, all photon energies are expressed in dimensionless units of $m_ec^2$. Following \citetalias{Wadiasingh2018}, we equate the final photon OF emission angle, $\mu_f $, with $\mu_B$, the angle between an observer at angle $\theta_v$ to the dipole axis and the polar angle of the local field direction. Assuming that the incident photon angle in the ERF is along the local field direction, a good approximation for a relativistic electron with Lorentz factor $\gamma\gg 1$, the initial ERF photon energy is determined by the scattering kinematics to be 
\be \label{eqn:what}
\omega_i = \hat\omega_i = {\omega_f(2 - \omega_f - \omega_f\cos^2\theta_f) \over 2(1 - \omega_f + \omega_f\cos\theta_f)}
\ee
where $\theta_f$ is the final photon angle in the ERF. The Jacobian $|\partial \omega_i / \partial \omega_f |$ in Eqn.~(\ref{eqn:ngamma}) is derived using Equation \ref{eqn:what}.   $n_\gamma(\varepsilon_s)$ is the soft photon density distribution, assumed to be a black body with emission over the entire NS surface, and $f(\mu_i)$ is the soft photon angular distribution at the position of the electron that incorporates the decrease in photon density with distance from the NS surface.  The final integral in Equation (\ref{eqn:ngamma}) is evaluated following the Appendix of \citetalias{Wadiasingh2018} \citep[see also][for details]{Baring2011}.  The scattering cross section $\sigma^{\perp,\parallel}$ for scattered  photons with $\parallel$ or $\perp$ polarization is adapted from \citet{Gonthier2014} and given in Equations (23) and (25) of \citetalias{Wadiasingh2018}; it is evaluated in the ERF.  The cross section is independent of the polarization of the incident photon since they are assumed to travel parallel to the local magnetic field in the ERF.  It treats incident photons in the cyclotron resonance and outside the resonance, taking into account the spin-dependence of cyclotron resonance \citep{2005ApJ...630..430B}.  

The only other integration in Equation~(\ref{eqn:ngamma}) is along the magnetic field loop, $s$.  \citetalias{Wadiasingh2018} normalize the scattered photon spectra at each observer angle to the total distance along the field loop in their template spectra, since they assumed the electrons maintain a constant Lorentz factor around the loop.  Since we will be allowing particles to lose energy as they radiate RICS emission, we follow a different protocol.  We compute the RICS emission spectrum from each step of the particle trajectory, assuming all the emission is approximately in a direction parallel to the local magnetic field.  To perform the $s$ integration at each step around the loop, we find the locations of the resonant points for each final photon energy from the condition $\hat\omega_i = B'$, where $B'$ is the local magnetic field (Equation (32) of \citetalias{Wadiasingh2018}), and integrate over fine steps in $ds$ around those locations.  To do this, we first find the polar angles corresponding to these points from the approximations given in Equations (34) - (37) of \citetalias{Wadiasingh2018} and then convert these to positions on the loop.  The spectra are then normalized to the sum of all the small integration steps, $\Delta S(\epsilon_f) \ll \Delta S_0$, where $\Delta S_0$ is the step length of the particle trajectory.  

To test our calculation of the scattered photon RICS spectra, we compute the spectra over the whole loop for a constant particle Lorentz factor, normalizing over the total loop length and compare with the spectra for different electron Lorentz factors and observer angles shown in Figures 6, 8 and 9 of \citetalias{Wadiasingh2018} (see examples in the Appendix).  Except for the difference in normalization, our spectra agree well for the highest several decades of scattered photon energy, but not as well at the lowest energies.  This is expected since we have assumed that the whole spectrum is emitted along the local magnetic field line from the loop position at each step, whereas this is not true of the photons at the lowest energies in the spectrum, which are emitted over a larger set of loop positions and wider angles of a broader ``Lorentz cone'' about $\hat{\boldsymbol{B}}$. These low-energy photons can also result from large-angle resonant scatterings (Wadiasingh et al. 2025).  However, since only the highest energy RICS photons are involved in precipitating cascades, the accuracy of the lower end of their spectra is not important in influencing cascade genesis.

The RICS energy loss rate is computed by integrating over the energy-weighted photon spectrum from each step
\be \label{eqn:Gdot}
\dot\gamma_{\rm RICS} =  {1 \over \gamma^2} \int_{\varepsilon_{f,min}}^{\varepsilon_{f,max}}  d\varepsilon_f \left({dn_\gamma^{\parallel} \over dtd\varepsilon_f} + {dn_\gamma^{\perp} \over dtd\varepsilon_f}\right) \epsilon_f 
\ee 
where the $1/\gamma^2$ factor approximates an integration over the scattered photon solid angle that is concentrated in the Lorentz cone.  The $\dot\gamma_{\rm RICS}$ values calculated in this way agree with those presented in Figure 10 of \citet{Baring2011}.  Although we initially choose a step length $\Delta S_0$ for the particle trajectory, in some cases the RICS loss rates are so large that $\Delta S_{\rm RICS} = c\gamma/\dot\gamma_{\rm RICS} < \Delta S_0$, in which case we limit those step lengths to a fraction of $\Delta S_{\rm RICS}$.

\subsection{Pair Production}  \label{sec:pairprod}

In a strong magnetic field a single photon can produce an electron-positron pair since the perpendicular momentum is absorbed by the field and only parallel momentum is conserved.  If photons are emitted nearly along the local magnetic field, as they typically are when the radiating particles are relativistic, the photons must travel some distance to sample the field line curvature and acquire the angles to put them above the threshold, $2mc^2/\sin\theta_{\rm kB}$, where $\theta_{\rm kB}$ is the angle between the photon momentum and local field.  The one-photon pair attenuation coefficient however, rises exponentially with both photon energy and $\vert \boldsymbol{B}\vert$ \citep[][see also Eq.~(\ref{eq:Tpp_exp_form}) below]{Erber-1966-RvMP,Daugherty1983}, and in pulsar-strength fields, photons must comfortably exceed the threshold before the probability for producing a pair becomes high enough.  Pairs are thus produced in excited Landau states and decay through cyclotron/synchrotron emission.  

Two-photon pair production is not likely to occur above one-photon pair threshold, but it is possible below that threshold.  However, the production rate will depend on the soft photon density and photon energies.  \citet{Burns1984} compared rates of one-photon and two-photon pair production in magnetized plasmas and found that the one-photon rate will exceed that of the two-photon rate for magnetar fields and surface temperatures unless the photon density exceeds $\sim 10^{28}\,{\rm cm^{-3}}$.  Even though those estimates were made for photon distributions and non-magnetic two-photon cross sections, the probability of two-photon pair production below one-photon pair threshold will be negligible for the photon densities we consider, namely $\lsim 10^{21}\,{\rm cm^{-3}}$ for a thermal distribution at $0.5$ keV, and much lower at $\lsim 10^{16}\,{\rm cm^{-3}}$ for standard RICS models of hard X-ray tail emission \citep{BaringHarding2007}.

In magnetar-strength fields, the probability for producing a pair is already high as soon as the photon reaches threshold, and so both the electron and positron are produced in the ground state in the case of $\parallel$-mode photons and are produced with one pair member in the first-excited state in the case of $\perp$-mode photons \citep{BaringHarding2001}.  Then, only $\perp$-mode photons can emit cyclotron photons.  Due to the near-threshold pair creation, pair production attenuation in magnetar fields must be treated more carefully.  

Therefore we follow the procedure of \citet{BaringHarding2001,Harding1997} and use the exact pair attenuation coefficients for pair creation in low-lying Landau states when 
$2 \omega' \leq 1 + (1+ 4B')^{1/2}$, where $\omega' = \varepsilon / (2\sin\theta_{\rm kB})$ and $B' = B/B_{\rm cr}$.  Unlike those of \citet{Daugherty1983}, they are derived using Sokolov-Ternov wavefunctions that give the correct  polarization-dependent  coefficients \citep{Sina1996}.  In the case of pairs produced in the ground state, the pair energy and momentum are assigned to Landau states with quantum numbers $(j,k) = (0,0)$ for $\parallel$ mode photons and $(j,k) = (0,1)$ or $(1,0)$ for $\perp$ mode photons.  The energy and momentum of the pair are
\be
E_j = (1 + p_{jk}^2 + 2jB')^{1/2},~~~       
E_k = (1 + p_{jk}^2 + 2kB')^{1/2}
\ee
with the pair having equal and opposite momentum,
\be
p_{jk} = \left[ {\varepsilon^2\sin^2\theta_{\rm kB} \over 4} - 1 - (j+k)B' + \left[{(j-k) B' \over \varepsilon\sin\theta_{\rm kB}}\right]^2 \right]^{1/2},
\ee
in the center-of-momentum frame.
For photons with energies higher above threshold, we use the asymptotic attenuation coefficients, corrected for near-threshold behavior \citep{Daugherty1983}
\be
T^{\rm pp}_{\parallel,\perp} = {\alpha \over 2\lambar} \, B'\sin\theta_{\rm kB} \, 
\exp \biggl( - {{4}\over {3\chi}} \biggr)
 \label{eq:Tpp_exp_form}
\ee
where $\alpha$ is the fine structure constant, $\lambar = \hbar/m_ec$ is the reduced electron Compton wavelength,  $\chi = \varepsilon B'\sin\theta_{\rm kB}/(2\,F)$ with $F = 1 + 0.42(\varepsilon\sin\theta_{\rm kB}/2)^{-2.7}$.  In this case, the photon energy is split equally among the pair and the pair Landau states are assigned using the formula \citep{Daugherty1983}
\be
j = k = {\varepsilon^2\sin^2\theta_{\rm kB} - 1 \over 2 B'} \,.
\ee
Although the pairs that are produced in excited states can have their spins up or down, the pair attenuation coefficients we use have been summed over the final spin state.  Therefore, we do not take electron or positron spins into account, even though their SR is dependent on spin state (see Section \ref{sec:SR}); treatment of this nuance is deferred to future work.

\subsection{Photon Splitting}  \label{sec:split}

Photon splitting, $\gamma \rightarrow \gamma\gamma$, is a QED process in which a single photon splits into two lower-energy photons.  It can take place efficiently only in strong magnetic fields $B' \gtrsim 1$, and unlike one-photon pair production, it does not have a threshold.  Therefore, even though it is a third order process in $\alpha$ and pair production is first-order in $\alpha$, it is possible for photons to split before reaching pair threshold when traversing the curved magnetic fields of the magnetosphere.  The original study by \citet{Adler1971} determined that the only kinematically allowed mode for splitting is $\perp \rightarrow \parallel\parallel$ in the limit of weak dispersion caused by the magnetized quantum vacuum.  However, there are three modes permitted by CP invariance: $\perp \rightarrow \parallel\parallel$, $\perp \rightarrow \perp\perp$, and $\parallel \rightarrow \perp\parallel$ and if the limit of weak dispersion does not hold in magnetar-strength fields, all three of these modes may be operating.  In that case, one-photon pair creation can be completely suppressed and a photon splitting cascade will ensue with no pairs or synchrotron emission produced \citep{Baring1998,Harding1997}.  Since there is presently no consensus on whether all three of these modes are operating in magnetar magnetospheres that are heavily loaded with plasma, we will assume only the $\perp \rightarrow \parallel\parallel$ splitting mode for this study.  Examples of splitting attenuation with all three modes are treated by \citet{BaringHarding2001} for general photon opacity considerations, and specifically for RICS signals by Wadiasingh et al. (2025).

The photon splitting rate for the $\perp \rightarrow \parallel\parallel$ mode is \citep{Adler1971}
\be
T^{sp}_{\perp \rightarrow \parallel\parallel} (\varepsilon) = {\alpha^3 \over 60\pi^2 \lambar}\,B'^6\,\varepsilon^5\sin^6\theta_{\rm kB}\, \bigl[ M(B') \bigr]^2
\ee
where $\varepsilon$ is in units of $m_ec^2$, as always, and
\be
M (B') = {1 \over { \bigl[ B'\bigr]^4}}\,\int_0^{\infty} \,{ds \over s}\,\exp(-s/B')\,\left[\left(-{3 \over 4s} + {s \over 6}\right){\cosh s \over \sinh s} + {3 + 2s^2 \over 12 \sinh^2 s} + 
{s \cosh s \over 2 \sinh^3 s} \right].
\ee
The integrals, $M (B')$, are computed and stored in tables for interpolation by the MC simulation. The energies of the split photons are determined from the differential splitting rate
\be \label{eqn:SPenergy}
T^{sp}_{\perp \rightarrow \parallel\parallel} (\varepsilon, \varepsilon_1) \simeq 30 {\varepsilon_1^2(\varepsilon - \varepsilon_1)^2 \over \varepsilon^5}\,
T^{sp}_{\perp \rightarrow \parallel\parallel} (\varepsilon),
\ee
valid in the limit of weak vacuum dispersion, $\varepsilon B'\sin\theta_{\rm kB} \lsim 1$, or below pair threshold \citep{Harding1997}.
The energy of one of the split photons ($\varepsilon_1$) is sampled from this distribution and the energy of the other photon is determined by the difference $\varepsilon_2 = \varepsilon - \varepsilon_1$ since both final photon momenta are co-aligned with that of the parent photon in the limit of weak vacuum dispersion.  

\subsection{Synchrotron Radiation in Strong Magnetic Fields}  \label{sec:SR}

In magnetar fields, synchrotron radiation (SR) is treated as the decay of electrons or positrons from excited Landau states.  To incorporate the discreteness of this process, we make use of the QED radiation rates from \citet{SokolovTernov1968}.  Although the SR rates are dependent on electron spin state, we do not take this into account since the pair attenuation rates that would determine the electron and positron spins have been averaged over spin. It is important to use QED emission rates when $\gamma B' > 0.1$ since in this regime, the classical formulae violate energy conservation (with emitted photons exceeding the electron energy).  In addition, QED-specific behavior such as the dominance of ground-state transitions, where electrons preferentially emit a single photon with nearly all its energy, is not captured in a classical treatment \citep{HardingPreece1987}.  For pairs in lower Landau states, $n < 20$, we use the exact cyclotron emission rates and for pairs in higher states we use the asymptotic formula \citep{SokolovTernov1968}, see also the Appendix of \citet{HardingPreece1987}.  The rates for cyclotron emission in the ``circular" frame where the electron momentum is perpendicular to B are computed and stored in tables as functions of local magnetic field and electron initial and final Landau state and final photon polarization for interpolation by the MC.  Likewise, the asymptotic synchrotron emission rates are stored in tables for local magnetic field, initial electron energy and Landau state and final photon energy and polarization.  These rates are time-dilated in boosting from the circular frame to the magnetospheric rest frame.  We assume that the photons are emitted in the  electron orbital plane (perpendicular to the magnetic field in the circular frame), which is a good approximation for electrons with relativistic energies in the lab frame.  The emitted photon azimuthal directions are assigned randomly.  After emission, the photon energy and direction are transformed to the OF by applying standard aberration formulae for the Lorentz factor of the radiating electron.  

\subsection{Description of Monte-Carlo Code}  \label{sec:MC}

\begin{figure} 
\centerline{\includegraphics
[width=11cm]{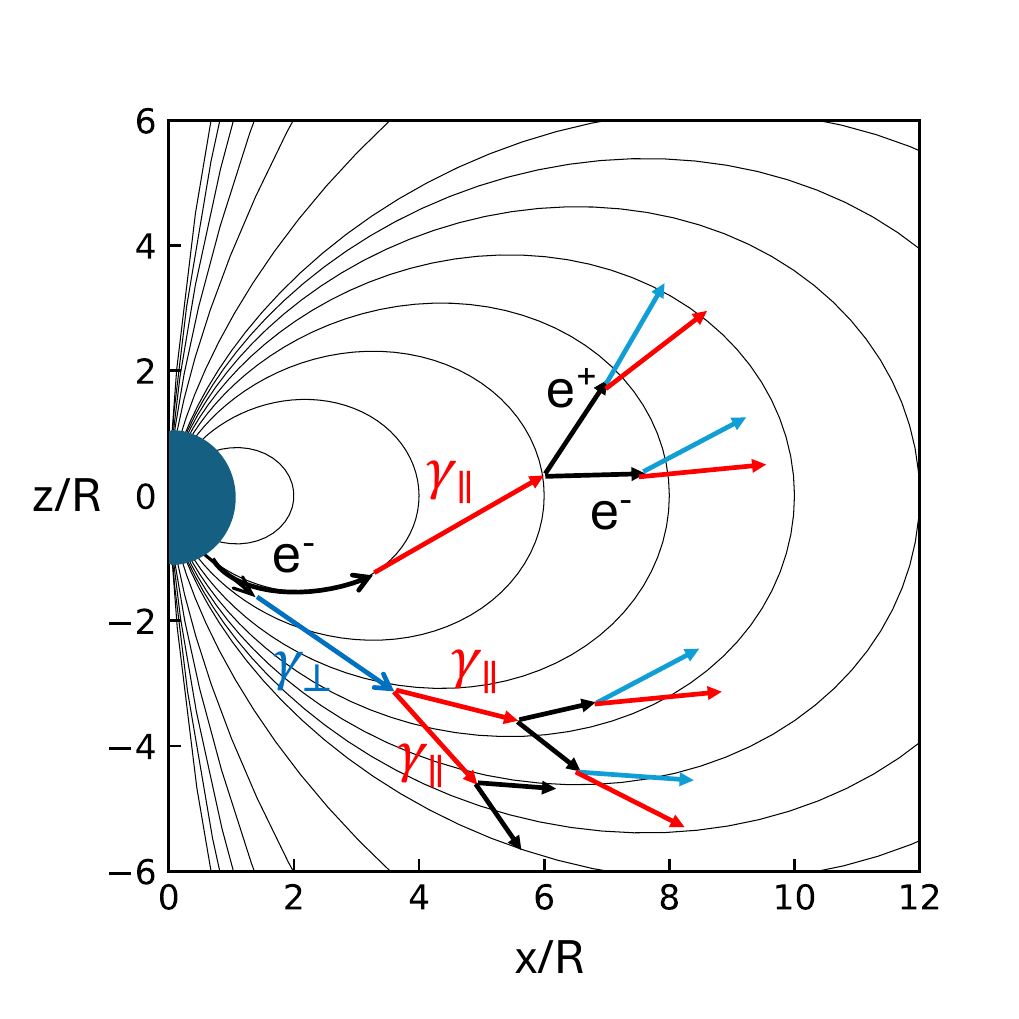}}
\vspace{-10pt}
\caption{Schematic of a pair cascade initiated by primary electrons traveling along a closed dipole meridional field loop, emitting resonant inverse Compton scattered photons with both $\parallel$ (red) and $\perp$ (blue) polarization.  Photon and pair trajectory segments are not to scale.}
\label{fig:cascade}
\end{figure}

We assume a single primary electron of initial Lorentz factor $\gamma_0$ travels around one closed field loop of a flat spacetime dipole field, defined by the maximum radius, $r_{\rm max}$ of the loop in units of NS radii, R, as shown in Figure \ref{fig:cascade}.  The loop is in the dipole x-z plane, the $\phi = 0$ meridional plane.  The electron emits RICS as described in Section \ref{sec:RICS} and the spectrum from each step is divided into a number of energy bins.  A single photon having the average energy of each bin, weighted by the number of RICS photons in that bin, is followed in a straight line path in the direction of the radiating electron (the tangent to the local field at that point).  In the present calculation, we ignore GR effects which can affect the photon mean free paths \citep{Harding1997,Story2014}.  As the photon propagates through the curved magnetic field, the mean free paths for pair production (pp) and photon splitting (sp) are accumulated independently
\be
\tau_{\rm pp,sp}(s) = \int_0^s\,T_{\rm pp,sp}(\varepsilon, \theta_{\rm kB})ds',
\ee
that determine the Poisson probability of whether the photon pair produces, splits or escapes,
\be
P^{\rm pp,sp}_{\rm surv} (s) = \exp[-\tau_{\rm pp,sp}(s)],
\ee
where $s$ is the distance along the photon path from its emission point.   For the initial RICS photons, the photon pair produces or splits when $P^{\rm pp,sp}_{\rm surv} < 1/e$, since only a single photon represents all photons in that energy bin.  For the higher generations of split photons or synchrotron photons, their outcome is determined as follows: when the product of the splitting and pair production survival probabilities, $P_{\rm surv}^{\rm sp}P_{\rm surv}^{\rm pp} > {\cal R}_1$, where ${\cal R}_1$ is a random number between 0 and 1, the photon escapes.  If not, then the probability that the photon survives splitting but not pair production is ${\cal P}_{\rm surv} = P_{\rm surv}^{\rm sp}(1 - P_{\rm surv}^{\rm pp})/(1-P_{\rm surv}^{\rm sp}P_{\rm surv}^{\rm pp})$. If ${\cal P_{\rm surv}} > {\cal R}_2$, where ${\cal R}_2$ is a second random number, then the photon pair produces; otherwise the photon splits.  

When the photon produces a pair, each member of the pair emits a series of cyclotron or synchrotron photons (if that particle is not initially in the ground Landau state) until it decays to the ground state. Since the SR decay rates are much higher than the RICS loss rates, we assume that all of the SR from each pair is emitted at its production point.  We do not follow any RICS radiation from the pairs, as their parallel Lorentz factors are generally low and the multiplicities are not high for cases when RICS energy losses are included, as will be discussed in Section \ref{sec:results}.  However, for cases when RICS energy losses are not applied, the multiplicity is much higher and RICS from pairs, and the cascades from that radiation, would be important.  When the photon splits, the daughter photons are assumed to have the same direction and energies chosen by Equation \ref{eqn:SPenergy} above.  Each synchrotron and split photon is followed to determine if it pair produces, splits or escapes, in the same way as for the RICS photons through a call to a recursive procedure that can follow an arbitrary number of photon generations.  The weights from the initial RICS photons are preserved through all the cascade generations.  Depending on the local field strength, photons of either polarization mode can pair produce or only $\parallel$ mode photons pair produce if splitting dominates near the NS surface.  The cascade from each primary particle step develops until all photons escape the magnetosphere.   In practice, escape occurs whenever either the photon radius is greater than $10^8$ cm or the local magnetic field is $B < 10^{-4}\,B_{\rm cr}$.

All photons are accumulated in energy, polarization, viewing angle and radiation type (RICS, synchrotron, split), with unattenuated and attenuated spectra accumulated separately.  All pairs are accumulated in an array that specifies three quantities: the final ground state (parallel) energy after SR, and the radius and the dipole polar angle that signify their position in the magnetosphere at their creation.  We also accumulate the sequences of pair production events and photon splittings in each pair generation.

\section{Results}  \label{sec:results}

The primary electron starts at the NS surface, having initial Lorentz factor $\gamma_0$, on the south-pole footpoint of a closed dipole field loop of maximum radial extent $r_{\rm max}$.  The starting polar angle is therefore $\theta_0 = \pi - \arcsin[(1/r_{\rm max})^{1/2}]$.  We assume blackbody radiation from the whole NS surface with temperature, $T = 5 \times 10^6$ K, typical of observed persistent thermal emission from magnetars.  The models are parametrized by the values of $T$, $\gamma_0$, $r_{\rm max}$, and surface magnetic field strength, $B_0$.  

\begin{figure} 
\includegraphics[width=140mm]{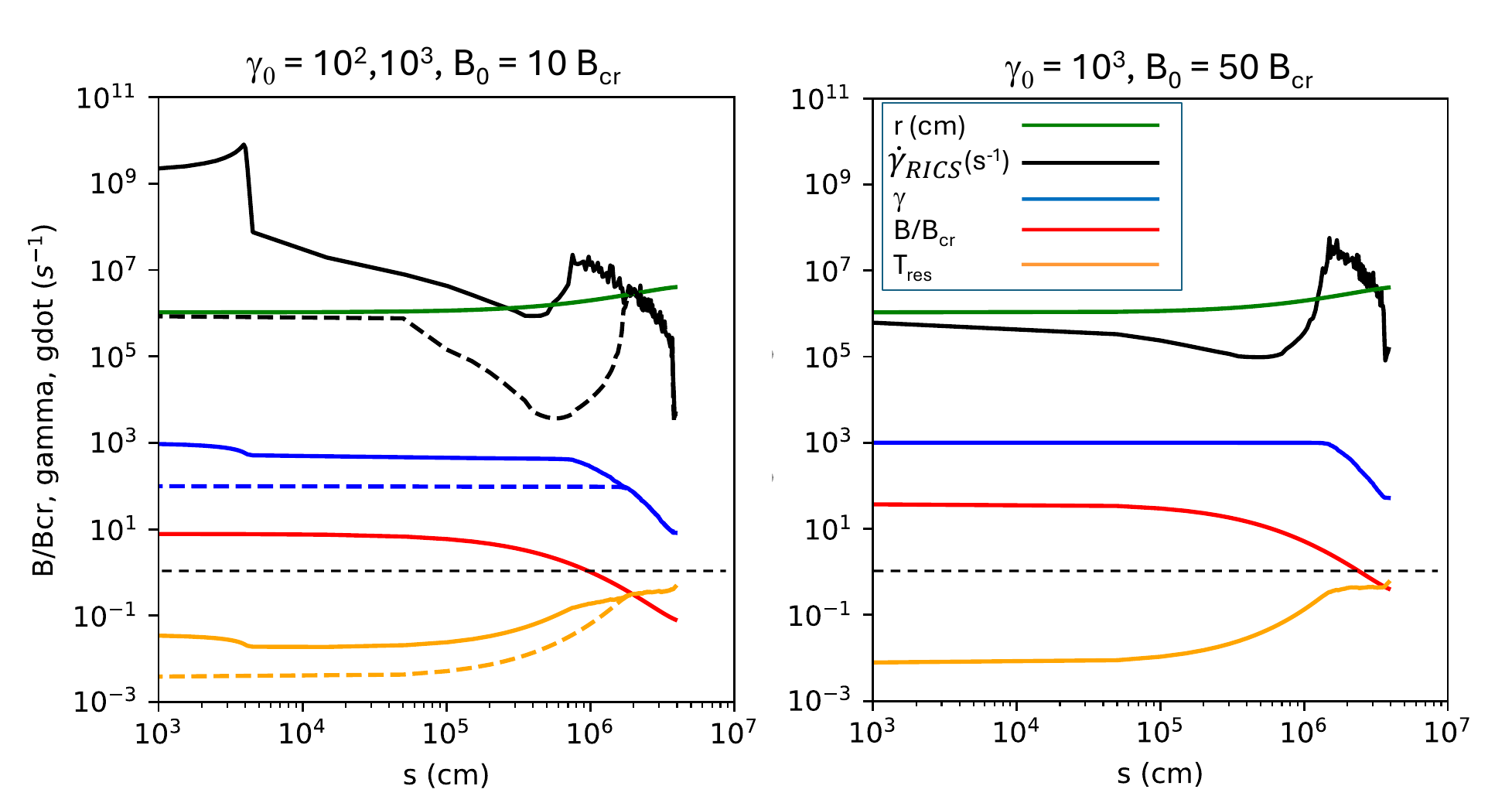}
\caption{Primary electron radius, $r$, resonant inverse Compton scattering loss rate, $\dot\gamma_{\rm RICS}$, Lorentz factor, $\gamma$, local magnetic field strength, $B/B_{\rm cr}$ and resonant condition, $T_{\rm res}$, as a function of distance $s$ along a closed dipole field loop with maximum radius $r_{\rm max} = 4$.  Left panel: $B_0 = 10\,B_{\rm cr}$ and $\gamma_0 = 10^3$ (solid lines), $\gamma_0 = 10^2$ (dashed lines). Right panel: $B_0 = 50\,B_{\rm cr}$ and $\gamma_0 = 10^3$.}
\label{fig:dynamics}
\end{figure}

\subsection{Particle Dynamics} \label{sec:dyn}

In the case where the primary electron loses energy to RICS, we use Equation (\ref{eqn:Gdot}) to reduce the Lorentz factor at each step along the trajectory.  As described above, the step length is limited so that the electron does not lose more than a fixed fraction (5\%) of its energy.  At the start of the trajectory, the radiation may be in the extreme Klein-Nishina regime where there is some probability that the particle could lose all its energy by emitting a single high-energy photon.  We do not capture these events because of the step loss limitation.  Additionally, we are not treating the RICS as a QED process (even though we use a QED cross section).  The RICS energy loss rate is very sensitive to local values of $\gamma$, $B$ and radius.  Figure~\ref{fig:dynamics} shows the RICS loss rate, $\dot\gamma_{\rm RICS}$, $\gamma$, local $B$ and radius as a function of distance along the trajectory, $s$ for two values of the surface magnetic field.  We also plot a dimensionless resonant condition $T_{\rm res} = 3\epsilon_0\gamma(1 - \beta\cos\theta_i)/B$, where $\epsilon_0 = kT/mc^2$ and $\theta_i$ is the angle between the soft photon direction, assumed to be the radial direction at most pertinent altitudes, and the particle velocity, assumed to be the local magnetic field direction.  Greater proximity of $T_{\rm res}$ to 1.0 (marked by the horizontal dashed lines in Figure~\ref{fig:dynamics}) at $r \gtrsim 10^6$cm then marks enhanced RICS loss rates above the stellar surface. In the case where $B_0 = 10\,B_{\rm cr}$ ($B_0$ is defined to be the field at the surface pole) and $\gamma_0 = 10^3$, $\dot\gamma_{\rm RICS}$ is very high since the particle is in resonance near the NS surface.  Its energy drops rapidly over a short distance until it is no longer in the resonance and $\dot\gamma_{\rm RICS}$ drops by several orders of magnitude.  The Lorentz factor then stays mostly constant as the particle coasts until the local field drops and the resonance condition is again satisfied, at which point the $\gamma$ decreases steadily.  In the case of $B_0 = 10\,B_{\rm cr}$ and $\gamma_0 = 10^2$, the electron is not in resonance at the NS surface, $\dot\gamma_{\rm RICS}$ is much lower and the Lorentz factor stays mostly constant until the local field drops and $\dot\gamma_{\rm RICS}$ rises.  The particle then begins to lose energy until the top of the loop.  In the case of a higher surface field, $B_0 = 50\,B_{\rm cr}$ and $\gamma_0 = 10^3$, the dynamics looks very similar to the $\gamma_0 = 10^2$ case in the left-hand panel, where the resonance is not reached until the local $B$ field drops.

\subsection{RICS Spectra}  \label{sec:RICS_spectra}

\begin{figure} 
\includegraphics[width=180mm]{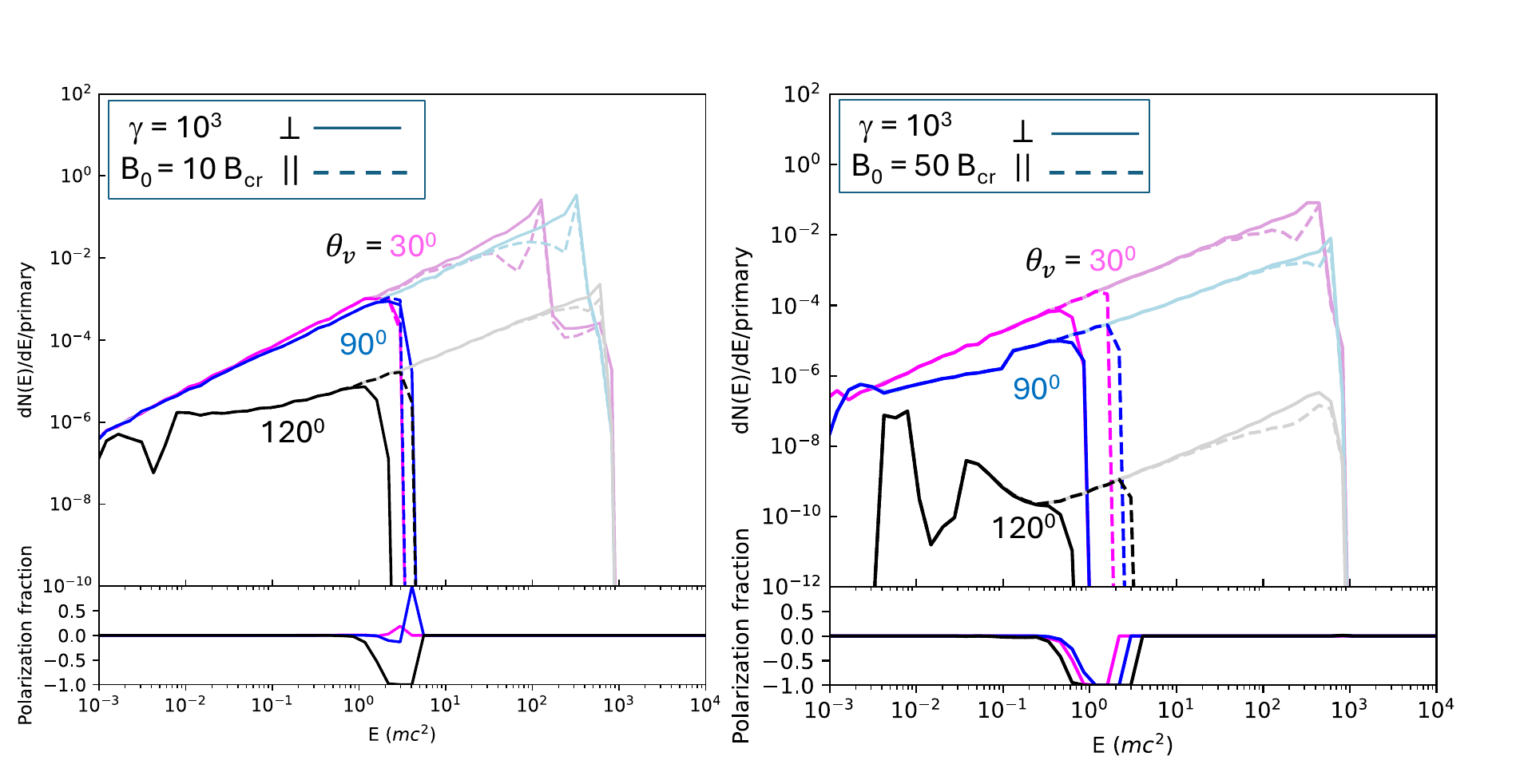}
\caption{Top panels: Resonant inverse Compton scattering photon spectra for $\parallel$ and $\perp$ photon polarization emitted by an electron with constant Lorentz factor $\gamma = 10^3$ on a closed dipole magnetic field loop with maximum radius, $r_{\rm max} = 4$, at different viewing angles to the dipole axis, as labeled, and for two different surface magnetic field strengths, $B_0$.  Light lines/dark lines are spectra unattenuated/attenuated by photon splitting and pair production. Bottom panels: Polarization fraction, $(\perp - \parallel)/(\perp + \parallel)$, as a function of photon energy for each of the above spectra.}
\label{fig:RICSg3nl}
\end{figure}

The RICS spectra at each step around the loop are computed from Equation (\ref{eqn:ngamma}) with the direction of the scattered photons determined at each point by the local magnetic field direction
\be \label{eqn:cthetav}
\cos\theta_v = -{3\cos^2\theta - 1 \over \sqrt{1 + 3\cos^2\theta}}
\ee
where $\theta$ is the polar angle of the electron position with respect to the dipole axis, defined to be the z-axis (see Figure \ref{fig:cascade}).  The negative sign on the right-hand side of Equation (\ref{eqn:cthetav}) is needed since the particle is moving opposite to the direction of the local field.  We normalize the RICS spectrum at each observer (polar angle) direction to the number of photons per energy interval emitted by one primary particle during the time it takes (number of steps) to radiate in that direction,
\be
{dn_\gamma^{\perp,\parallel}(\varepsilon_f, \theta_v) \over d\varepsilon_f} = {dn_\gamma^{\perp,\parallel}(\varepsilon_f, \theta_v) \over dtd\varepsilon_f}\,{\min(\Delta S_{\rm RICS}, \Delta S_0)\over c\gamma^2}
\ee
where the factor of $1/\gamma^2$ approximates the integral over final photon angle, and thereby captures the relativistic beaming of the emission.   The sum of this quantity over the final photon energy times $c/\min(\Delta S_{\rm RICS},\Delta S_0)$ is then equal to the RICS cooling rate over that step.  The total emitted spectrum radiated at each polar angle can then be obtained by multiplying the above spectrum by the total current of primary particles.

Figure \ref{fig:RICSg3nl} shows the RICS spectra emitted at different viewing angles, $\theta_v$, for two different surface magnetic field strengths, for a primary electron with no energy losses, so that the Lorentz factor is constant at $\gamma = \gamma_0 = 10^3$ around the loop.  This case with no RICS losses is not realistic since it would require the losses to be balanced by a constant acceleration.  We include the no-loss cases to benchmark the comparison with \citetalias{Wadiasingh2018} results, and also with the subsequent RICS loss cases we present here.  The total emitted (unattenuated) spectra are displayed as light lines and the spectra attenuated by pair production and photon splitting are displayed as dark lines; the treatment of cascade contributions to spectra is given in Section~\ref{sec:cas_spectra}.  The unattenuated RICS spectra are extremely flat, and the $\parallel$ and $\perp$ mode photons show significant differences only near the highest energies, where the $\perp$ mode dominates; this character is a direct consequence of the polarization of the cross section, as first described in \cite{BaringHarding2007}.  The attenuation cutoffs in the spectra are very polarization dependent since, as discussed in Section \ref{sec:model}, we assume that only $\perp$-mode photon split into two $\parallel$-mode photons.  The splitting cutoff can be much lower than the pair production cutoff of the $\parallel$-mode photons, depending on local field strength.  The lower panels in each plot shows the polarization fraction, $(\perp - \parallel)/(\perp + \parallel)$.  When photon splitting dominates, the spectrum near the cutoff can be 100\% polarized in $\parallel$ mode.  For these examples, this interestingly falls in the MeV range, and provides a motivation for future polarimetric capability in this band by planned missions such as COSI \citep{Tomsick-2024-icrc}
and AMEGO-X \citep{Caputo-2022-JATIS}.

\begin{figure} 
\includegraphics[width=180mm]{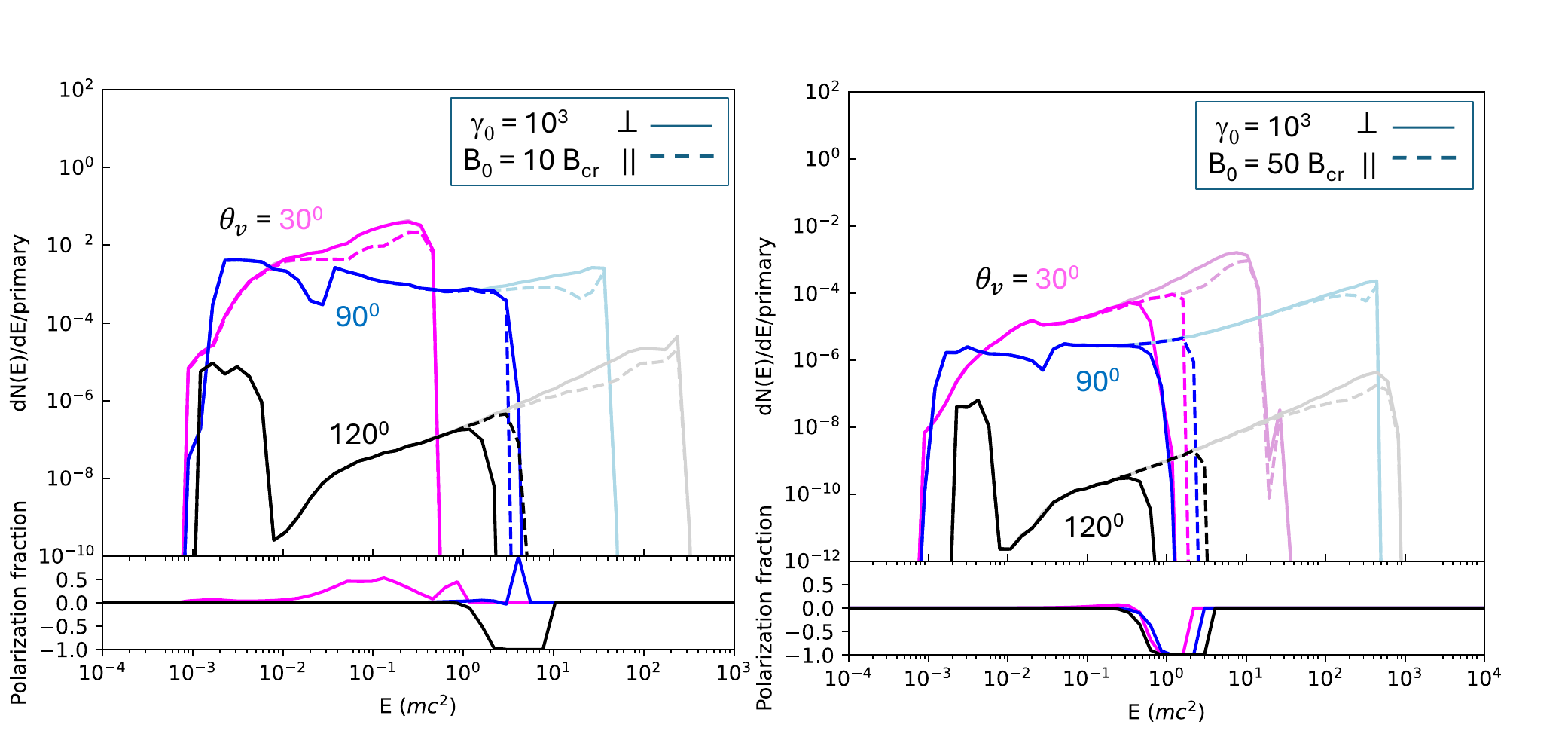}
\caption{Same as Figure \ref{fig:RICSg3nl} but with resonant Compton scattering energy loss included in the electron dynamics.}
\label{fig:RICSg3}
\end{figure}

The $\theta_v = 120^\circ$ spectrum is emitted  when the primary electron is still very near the NS surface (see Figure \ref{fig:cascade}).  Since the local magnetic field is very high at this point in the trajectory, photon splitting dominates over pair production and their attenuation cutoffs are quite distinguishable.  As a consequence, the spectrum just before the cutoff is 100\% polarized in $\parallel$ mode.  The pair attenuation cutoffs are at very similar photon energies for the two values of $B_0$ shown in Figure \ref{fig:RICSg3nl}, but the splitting cutoff is at a much lower energy for the $B_0 =50\,B_{\rm cr}$ case, reflecting the stronger dependence of the photon splitting attenuation energy on the field strength; the pair creation turnover energy couples more to its threshold.  The spectra for the smaller $\theta_v$ that are emitted at larger radii show very different cutoff behavior at the two different field strengths.  In the $B_0 =10\,B_{\rm cr}$ case, the local field has dropped below the strength where splitting dominates and pair production attenuates the spectrum for both photon polarization modes.  However, for $\theta_v = 90^\circ$, pairs are produced at threshold where pairs from $\parallel$-mode photons occupy the ground Landau state $(0,0)$ but $\perp$-mode photons must occupy the first-excited Landau state $(0,1)$ or $(1,0)$ with a higher threshold energy.   This produces a spike of 100\% $\perp$ mode at the cutoff. For $\theta_v = 30^\circ$, the photons are emitted near the top of the field loop, where the local field is even lower and pairs are produced further above threshold in excited states.  The pair cutoff for both modes is then at the same energy.  In the $B_0 =50\,B_{\rm cr}$ case, the spectra for all viewing angles are cutoff by splitting so that the $\perp$-mode cutoffs are all lower than the $\parallel$-mode cutoffs.  But notice that while the splitting cutoffs are at higher energies for the larger $\theta_v$, the pair cutoffs are at lower energies.   This is because the photons at the larger $\theta_v$ are emitted on local field lines with smaller curvature, producing smaller mean-free paths for pair production.

The RICS spectra for cases where the primary electron suffers radiative losses is shown in Figure \ref{fig:RICSg3} for two different surface magnetic field strengths.  The unattenuated spectra at lower viewing angles show progressively lower kinematic energy cutoffs as the particle loses energy.  In the $B_0 =10\,B_{\rm cr}$ case, with higher levels of RICS losses earlier in the trajectory (see Figure \ref{fig:dynamics}), the spectrum at $\theta_v = 30^\circ$ is even kinematically cutoff below the pair and splitting cutoffs.  Overall, the RICS cooling is greater for higher $\gamma$, and so acts to steepen the spectra relative to those in Figure~\ref{fig:RICSg3nl}. In the $B_0 =50\,B_{\rm cr}$ case, the electron loses energy at a much lower rate so that the unattenuated spectral cutoffs stay above the pair and splitting cutoffs and the spectra look similar to those without RICS losses.

\begin{figure} 
\includegraphics[width=180mm]{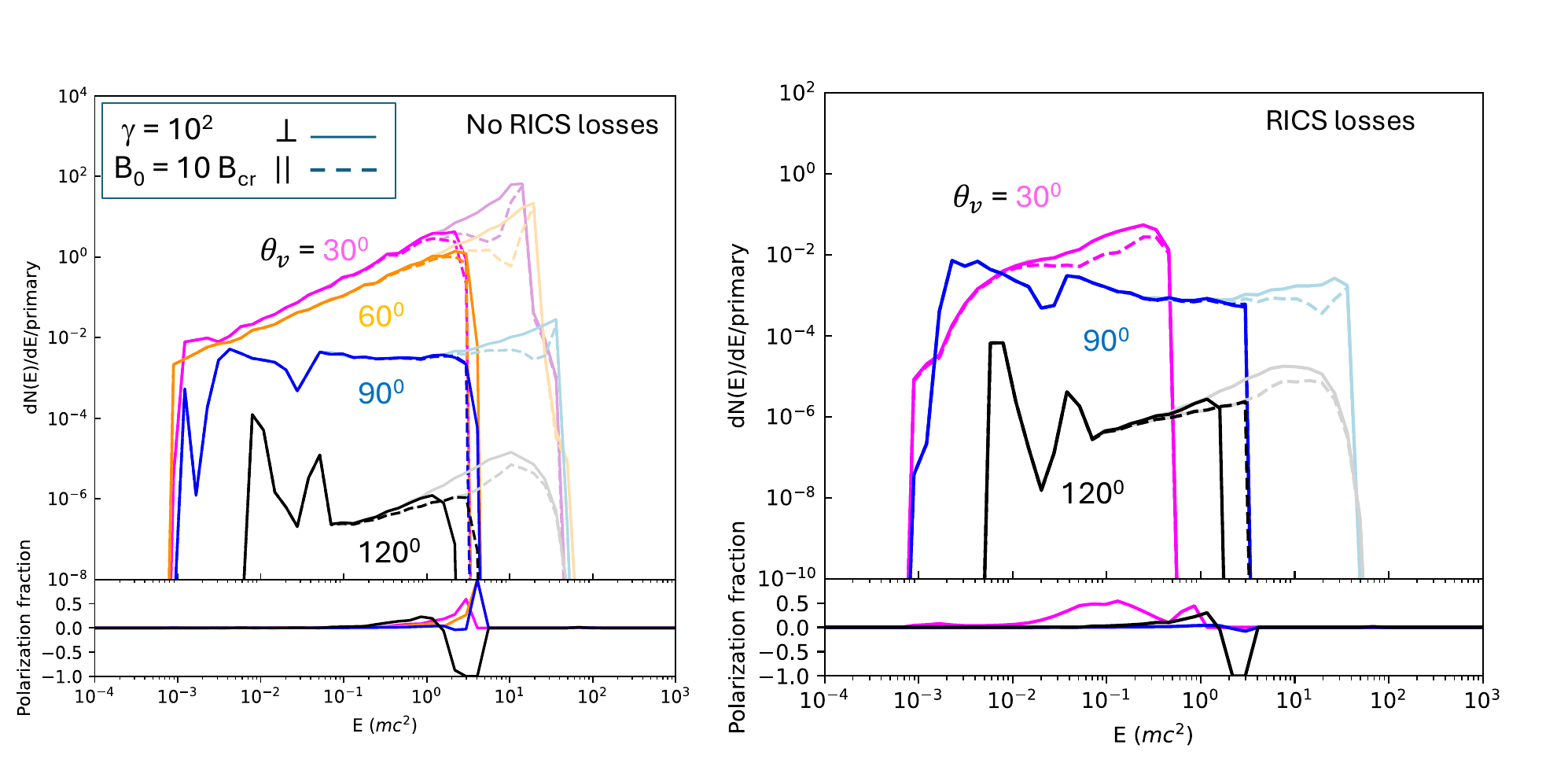}
\caption{Left panel: Same as Figure \ref{fig:RICSg3nl}, left panel, for electron Lorentz factor  $\gamma_0 = 10^2$ and surface dipole field strength, $B_0 =10\,B_{\rm cr}$.  Right panel: Same as left panel but with resonant Compton scattering energy loss included in the electron dynamics.}
\label{fig:RICSg2b10}
\end{figure}

Figure \ref{fig:RICSg2b10} shows RICS spectra for a lower initial electron Lorentz factor, $\gamma_0 = 10^2$, and $B_0 =10\,B_{\rm cr}$, for cases without and with RICS losses.  The unattenuated spectra have lower high-energy kinematic cutoffs, limited by the particle energy, but also have higher low-energy cutoffs since there is a reduced phase space for resonance interactions (see \citetalias{Wadiasingh2018}).  Since the pair creation and photon splitting opacities are not connected to RICS kinematics, the high-energy attenuation cutoffs are similar to those of the $\gamma_0 = 10^3$, $B_0 =10\,B_{\rm cr}$ case.  Interestingly, the spectra including RICS energy losses are very similar to those of the higher $\gamma_0 = 10^3$, $B_0 =10\,B_{\rm cr}$ case of Figure \ref{fig:RICSg3} at the larger viewing angles.  Even though the particle starts out with higher Lorentz factor, there is a saturation level near the top of the field loop which for $B_0 =10\,B_{\rm cr}$ is near $\gamma = 20$, a limit also noted by \citet{Baring2011} and \citet{Belo2013b}.  

\citet{Hu2019} presented pair production and photon splitting escape energies for photons emitted parallel to closed magnetic dipole field loops, both in flat spacetime, and also for the Schwarzschild metric.  We have compared our high-energy spectral cutoff energies with their examples in the absence of GR modifications and they agree very well.  GR effects will lower the escape energies and cutoffs due primarily to the higher effective surface magnetic field strength and photon energy in the local inertial frame near the NS surface \citep{Gonthier1994,Harding1997,Story2014}.

\subsection{Cascade Spectra}  \label{sec:cas_spectra}

\begin{figure} 
\includegraphics[width=180mm]{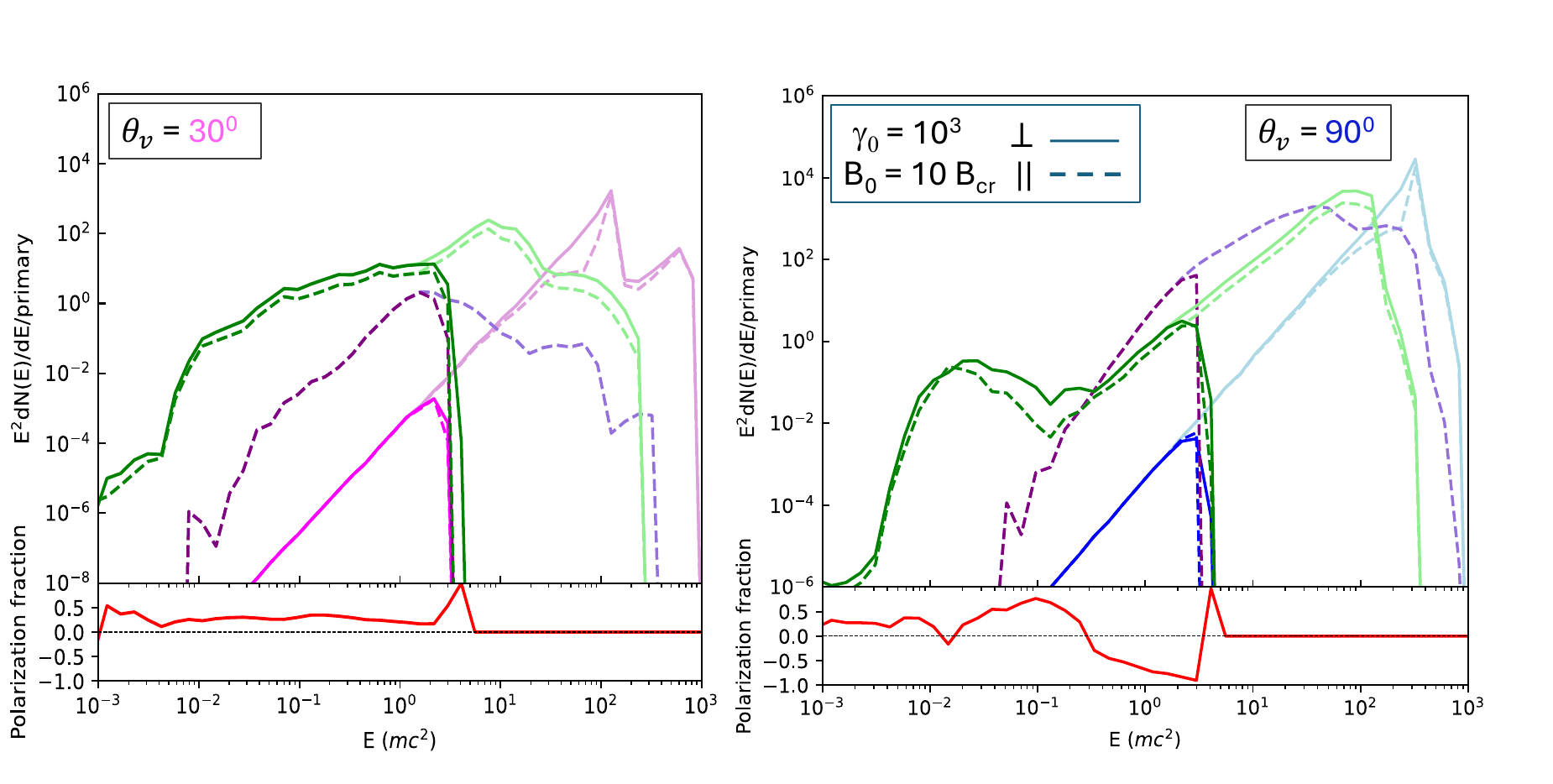}
\caption{Top panels: Photon spectral energy distributions of primary resonant inverse Compton scattering (pink, $\theta_v = 30^\circ$ and blue, $\theta_v = 90^\circ$), photon splitting (purple) and pair synchrotron radiation (green) from a pair cascade on a closed dipole field loop with maximum radius, $r_{\rm max} = 4$ for $\parallel$ and $\perp$ photon polarization from a primary electron with no RICS energy losses.  Bottom panels: Polarization fraction, $(\perp - \parallel)/(\perp + \parallel)$, as a function of photon energy for the total spectra above.}
\label{fig:caspecg3b10r4nl}
\end{figure}

The attenuation of the RICS spectra will result in additional spectral contributions from split photons and SR from pairs.  In many observer directions these contributions will overwhelm the RICS photon spectra, principally because the energy in secondary photons is derived from that of the attenuated primary gamma-rays above a few MeV.  Figure \ref{fig:caspecg3b10r4nl} shows the $\theta_v = 30^\circ$ and $\theta_v = 90^\circ$ RICS spectra from Figure \ref{fig:RICSg3nl} with the photon splitting and pair SR spectra, both of which are several orders of magnitude higher than the RICS spectrum.  The attenuated SR spectrum dominates at all energies for $\theta_v = 30^\circ$ and up to $\sim 0.1\,mc^2$ for $\theta_v = 90^\circ$, and all spectra are attenuated around $3\,mc^2$ by pair production.  The high-energy unattenuated SR spectra are radiated by pairs produced by the most energetic RICS $\parallel$-mode photons that are above threshold for creating pairs in low-lying excited states. For viewing angle $\theta_v = 30^\circ$, the attenuated SR spectrum with a photon index $\sim 1.5$ is much softer than the RICS spectrum as the pairs have a broad energy spectrum (see Section \ref{sec:pairs}), and is commensurate with observed indices of magnetar hard tails \citep{Kuiper-2004-ApJ,Kuiper-2006-ApJ,denHartog-2008-AandA,Enoto2010}.  The polarization is near 50\% in $\perp$ mode since SR dominates over the split photon spectrum.  For the $\theta_v = 90^\circ$ case, the SR spectrum still dominates at the lower energies, but the split photon spectrum begins to dominate above $1.0\,mc^2$ and the polarization switches from mostly $\perp$ to $\parallel$ mode until the cutoff.  In this case, pair production cuts off the spectra in both modes but at different energies.  The different pair production cutoff energies of the RICS spectra for $\perp$ and $\parallel$ modes were also seen in the left panels of Figures \ref{fig:RICSg3nl} and \ref{fig:RICSg3} and in Figure \ref{fig:RICSg2b10}, but now the split photon  and SR spectra are also cut off in the same way. Again, this produces a spike of 100\% $\perp$ mode around the cutoff energy.  The cascade spectra and polarization vary with energy in a manner that is very angle-dependent, since photon splitting is more important at lower radii (larger viewing angles) and SR is more important at larger radii (smaller viewing angles).

\begin{figure}[t] 
\includegraphics[width=180mm]{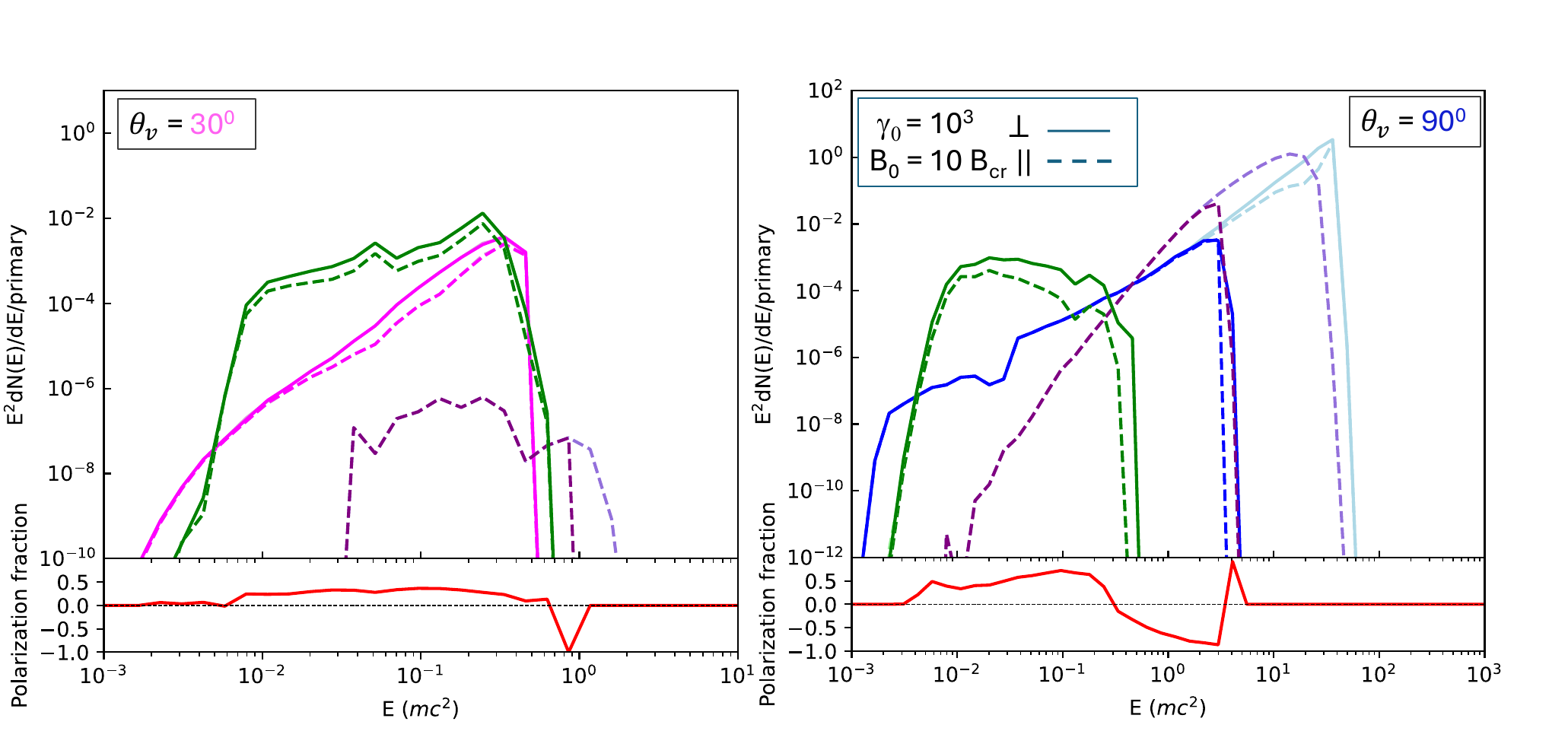}
\caption{Same as Figure \ref{fig:caspecg3b10r4nl} but with resonant Compton scattering energy loss included in the primary electron dynamics.}
\label{fig:caspecg3b10r4}
\end{figure}

\begin{figure}[b] 
\includegraphics[width=180mm]{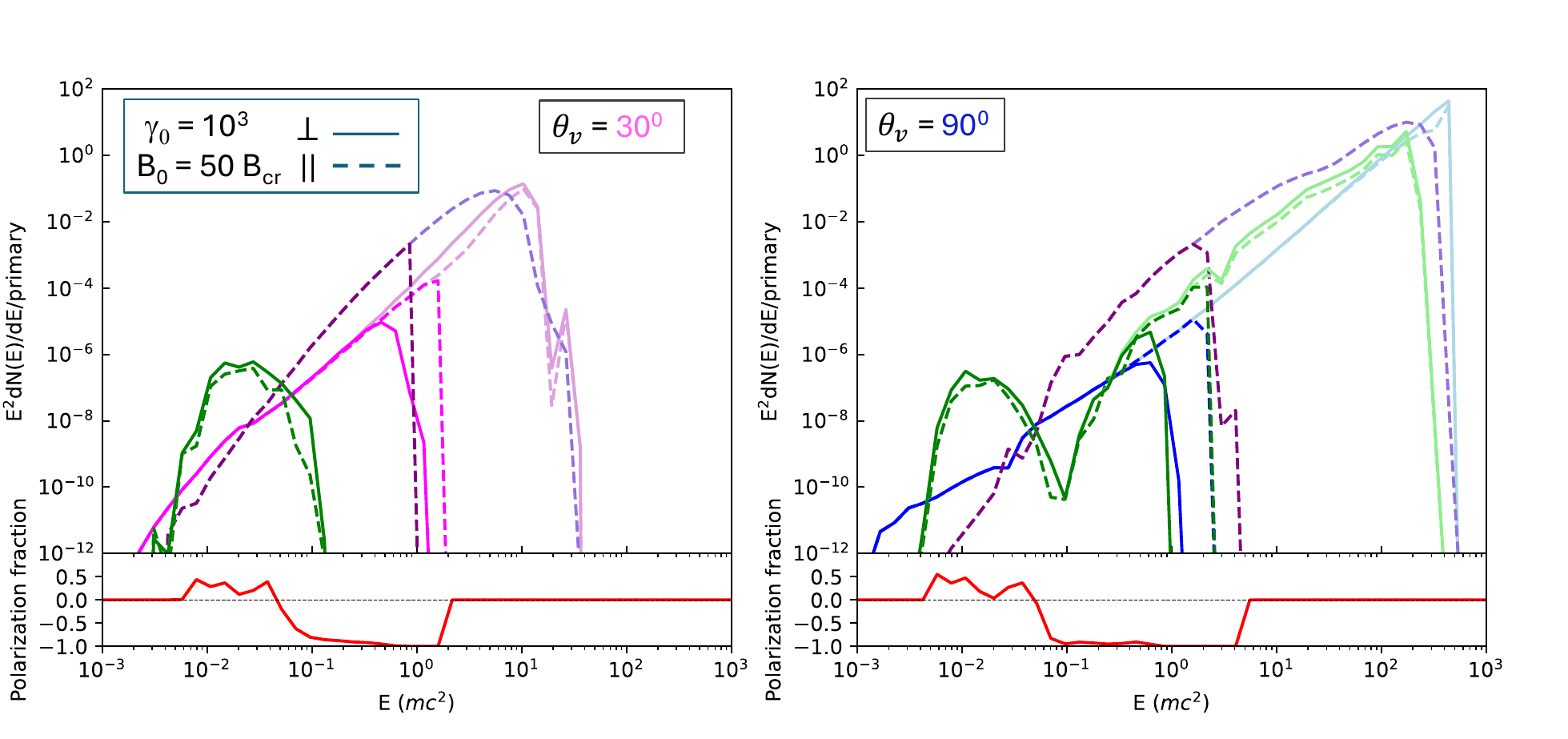}
\caption{Same as Figure \ref{fig:caspecg3b10r4} for surface magnetic fields strength $B_0 =50\,B_{\rm cr}$.}
\label{fig:caspecg3b50r4}
\end{figure}

Figure \ref{fig:caspecg3b10r4} shows the RICS spectrum for the viewing angles $\theta_v = 30^\circ$ and $\theta_v = 90^\circ$ and $\gamma_0 = 10^3$ in the case where RICS losses are applied, as well as the split photon and pair SR spectra.  For $\theta_v = 30^\circ$ the softer SR spectra with photon index $\sim 1.5$ again dominates at the lower photon energies starting around 3-4 keV but cuts off at $\sim 0.6mc^2$, below the pair production cutoff.  The much lower flux and lower cutoff energy compared to the no loss case is due to a much lower multiplicity and lower cutoff of the pair spectrum (see Section \ref{sec:pairs}). The split photon contribution is very small and causes only a reversal of polarization from $\perp$ to $\parallel$ near the SR cutoff.  The spectra for viewing angle $\theta_v = 90^\circ$, shown in Figure \ref{fig:caspecg3b10r4} are again dominated by the pair SR component at low photon energies, which is softer than in the $\theta_v = 30^\circ$ case, but cuts off sooner.  Above $\sim 0.4 mc^2$ the total spectrum is dominated by the RICS spectrum and by the split photon spectrum in a small spectral window.  The split photon spectrum, in pure $\parallel$ mode, is cutoff by pair production at the same energy as the $\parallel$ mode RICS spectrum and at a slightly lower energy than the RICS $\perp$ mode cutoff, due to the different pair thresholds of the two modes.  The total spectrum polarization with energy is therefore quite complex, with 40\% - 50\% $\perp$ mode at lowest energies transitioning through 0\% at RICS dominance, to nearly 100\% $\parallel$ mode as split photons dominate, ending with a spike of 100\% $\perp$ mode at the cutoff.  The cascade spectra for the same case for $\gamma_0 = 10^2$ look very similar to Figure \ref{fig:caspecg3b10r4} since the RICS spectra in the left panel of Figure \ref{fig:RICSg3} and the right panel of Figure \ref{fig:RICSg2b10} including electron energy losses are similar.  However, the lower kinematic cutoff energies of the unattenuated spectra in the $\gamma_0 = 10^2$ case yield fewer pairs (see Table 1).

The RICS and cascade spectra for $\gamma_0 = 10^3$  are shown for a higher field strength $B_0 =50\,B_{\rm cr}$ in Figure \ref{fig:caspecg3b50r4}.  In this case, pair SR is dominant over a much smaller part of the attenuated spectrum and the split photon spectrum dominates from $\sim 0.1mc^2$ up to the cutoff.  Since the split photon spectrum is purely $\parallel$ mode, it is attenuated by pair production. So the total spectrum polarization is in $\perp$ mode at the lowest energies but switches to nearly 100\% $\parallel$ mode until the cutoff for both viewing angles.

\subsection{Pair Distributions and Spectra} 
 \label{sec:pairs}

\begin{figure}[b] 
\hskip 2.0cm
\includegraphics[width=110mm]{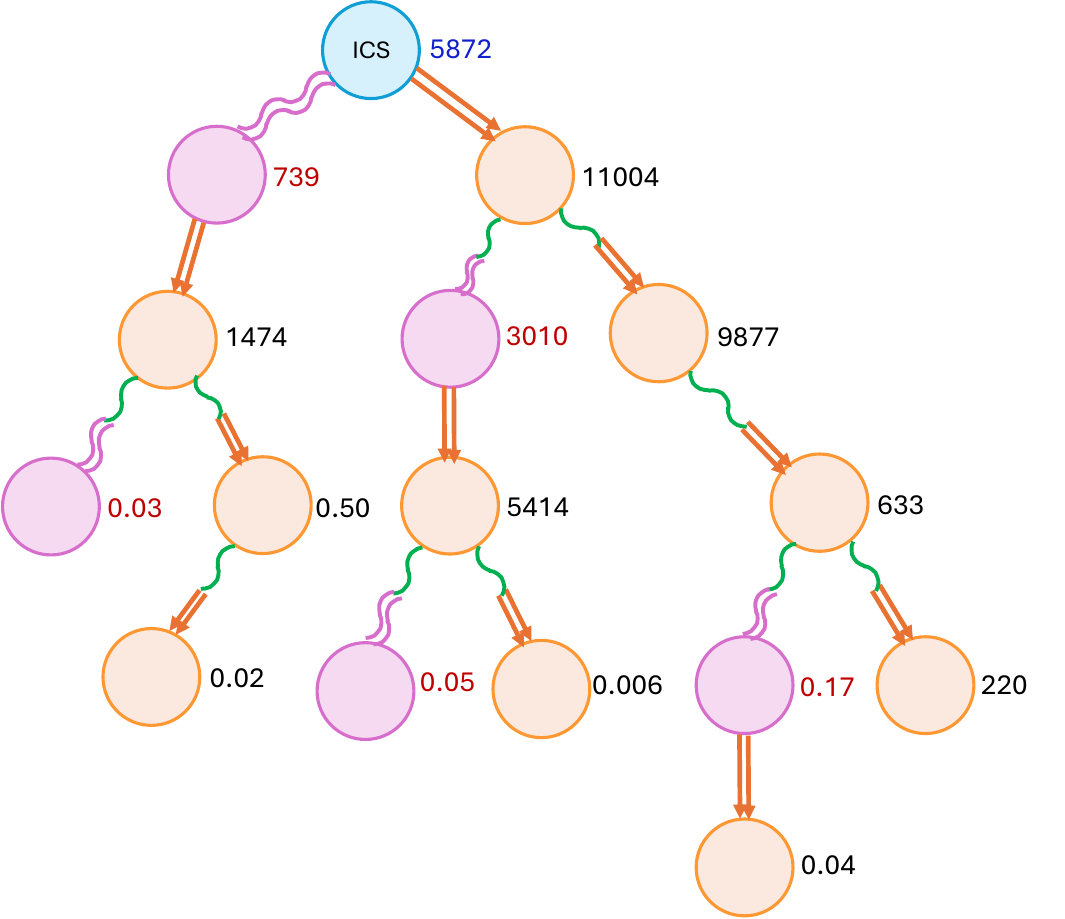}
\caption{Genealogy of cascade from a single primary electron of constant Lorentz factor, $\gamma_0 = 10^3$, on a closed field loop of maximum radius $r_{\rm max} = 4$, with surface magnetic field strength $B_0/B_{\rm cr} = 10$ (see also right panel of Figure \ref{fig:pairdistg3r4nl}).  Different generations, increasing from top to bottom occupy each row and the numbers of photons (blue for ICS photons and pink for split photons) and pairs (black) in each generation that are produced from the previous generation are shown.  Blue, pink and orange circles are ICS photons, split photons and pairs respectively.   Double orange arrows represent pair production.  Double pink squiggles represent photon splittings, with $\perp$ mode photons producing two $\parallel$ mode photons.  Green squiggles represent pair synchrotron photons that either split or pair produce.}
\label{fig:cascademat}
\end{figure}

The genealogy of a cascade from a single primary electron  with constant Lorentz factor that produces high multiplicity is shown in Figure \ref{fig:cascademat}.  It displays the processes that produce pairs and split photons at each generation of the cascade.  Generation 0 consists of RICS photons from all trajectory steps, a fraction of which produce a pair (orange  arrows) or split into two lower-energy photons (pink squiggles), making up generation 1.  The number of split photons and pairs in generation 1, divided by two, is nearly equal to the total number of RICS photons since the RICS spectrum is very hard and the photon number above the cutoffs dominates over the number below the cutoffs (see Figure \ref{fig:caspecg3b10r4nl}).  There are many more pairs than split photons in generation 1 since splitting only dominates in the first trajectory steps near the NS for this case with surface magnetic field strength $B_0/B_{\rm cr} = 10$.  For most other steps at higher altitude, pair production dominates and pairs are produced by photons in both polarizations modes.  In generation 1, the split photons from generation 0 produce pairs predominantly in the ground Landau state that do not radiate any synchrotron photons, stopping further significant cascading from that branch.  Many other pairs in generation 1 are produced in excited states by RICS photons emitted at several stellar radii from the NS, and they produce both split photons and pairs in excited states that radiate synchrotron photons.  The split photons in generation 2, produced at higher altitudes, can produce 3rd generation pairs above threshold in excited states.  There are effectively extremely few split photons in generation 3 and higher, primarily because at the associated high altitudes the magnetic field strength has dropped, and consequently the rest of the cascade consists of synchrotron photons producing pairs.

\begin{table}[h]
\caption{Pair Generations and Multiplicity}
\vskip 0.5 cm
\hskip -2.0cm
\begin{tabular}{c|c|c|ccc|ccc}
\hline\noalign{\smallskip}

\multicolumn{3}{c}{} & \multicolumn{3}{c}{No RICS losses} &  \multicolumn{3}{c}{RICS losses}  \\
\multicolumn{3}{c}{} & \multicolumn{3}{c}{$\gamma_0$} & \multicolumn{3}{c}{$\gamma_0$} \\
\hline
$B_0/B_{\rm cr}$ & $r_{\rm max}$ && $10$ & $10^2$ & $10^3$ & $10$ & $~~10^2~~$ & $~~10^3~~$ \\
\hline
10 & 4 & $N_{\rm gen}$ & 3  & 4 & 5 & 3 & 6 & 4 \\
   &  & Multiplicity & $9 \times 10^{-8}$ & $4.8 \times 10^4$ & $2.7 \times 10^4$ & $9 \times 10^{-8}$ & 15 & 62 \\
50 & 4 & $N_{\rm gen}$ & & 5 & 5 &  & 4 & 5 \\
   &  &  Multiplicity & & $7.8 \times 10^3$ & $7.8 \times 10^3$ & & 5 & 20 \\
\hline
10 & 6 & $N_{\rm gen}$ & & 6 & 6 & & &  \\
     & & Multiplicity & & $4 \times 10^4$ & $2.4 \times 10^4$ & & & \\

\hline
\end{tabular}
\end{table}

When RICS energy losses of the primary electrons are not included pair production is more abundant, since the RICS spectrum extends to much higher energies, and the cascades extend to five or six pair generations.  Table 1 shows the pair generations $N_{\rm gen}$, and multiplicities (number of pairs per primary electron) for different combinations of initial primary electron energy, surface magnetic field strength and maximum loop extent.  It is noteworthy that in the case of no RICS losses, the multiplicities are high and at levels comparable to those realized in curvature radiation-initiated polar cap cascade models \citep{HardingMuslimov2011,Timokhin2015} of GeV-band gamma-ray pulsars such as Vela.   The introduction of RICS losses reduces the maximum energy of super-MeV photons and thereby curtails the production of pairs dramatically.  Generally, the pair multiplicity decreases with $B_0/B_{\rm cr}$, since at the higher surface field strengths, photon splitting attenuation is dominant for the $\perp$ mode over a larger part of the primary electron trajectory.  

The dependence of pair multiplicity with $\gamma_0$ is more complicated.  In the no RICS loss case, the multiplicity increases dramatically from $\gamma_0 = 10$ to $\gamma_0 = 10^2$ and then saturates or decreases for $\gamma_0 =10^3$.  This behavior is a consequence of the RICS spectra, as illustrated by \citetalias{Wadiasingh2018} in the left-hand panel of Figure 6 for $B_0/B_{\rm cr} = 10$.  For $\gamma_0 = 10$, the particle is barely sampling the resonance and the maximum scattered photon energy lies below the attenuation cutoff, so one would not expect any pairs produced.  For $\gamma_0 = 10^2$ and $\gamma_0 = 10^3$, the resonance is sampled over a large range of scattered photon energies, but even though the maximum energy $\varepsilon_f \sim \gamma_0$ is higher for the higher $\gamma_0$, for most viewing angles the rate of photon production is lower by about two orders of magnitude, as also shown in the left-hand panels of Figures \ref{fig:RICSg3nl}, \ref{fig:RICSg2b10}, \ref{fig:compW2018} and Figure~6 of \citetalias{Wadiasingh2018}\footnote{Due to the $\gamma$ dependence in Equation (13) of \cite{BaringHarding2007}.}.   There are therefore more photons attenuated at lower energies in the $\gamma_0 = 10^2$ RICS spectrum, producing a prominent bump in the pair spectrum at lower energies (see Figure \ref{fig:pairspecNL}).  In the RICS loss case, due to the rapid RICS energy losses, the spectra and fluxes are not that different for the $\gamma_0 = 10^2$ and $\gamma_0 = 10^3$ cases and for most viewing angles, the number of attenuated photons is higher for higher $\gamma_0$.  However, the multiplicities for the RICS loss case are much lower since attenuation and pair cascading occurs over a smaller part of the trajectory, since the effective maximum energy of RICS photons is dramatically reduced by the strong RICS cooling.   The number of pair generations is highest for the larger $r_{\rm max} = 6$ case since the trajectory samples larger distances from the NS where the field strength is lower.  At the lower field strengths, pairs are produced in higher Landau states and generate more SR photons that produce more pairs.  The high generation number in the $\gamma_0 = 10^2$ case for no RICS losses also results from pairs being produced at larger distances from the NS, since the RICS photons have lower energies and longer pair production mean-free paths (see also Figure \ref{fig:pairdistg3r6nl}).

\begin{figure}[h] 
\includegraphics[width=180mm]{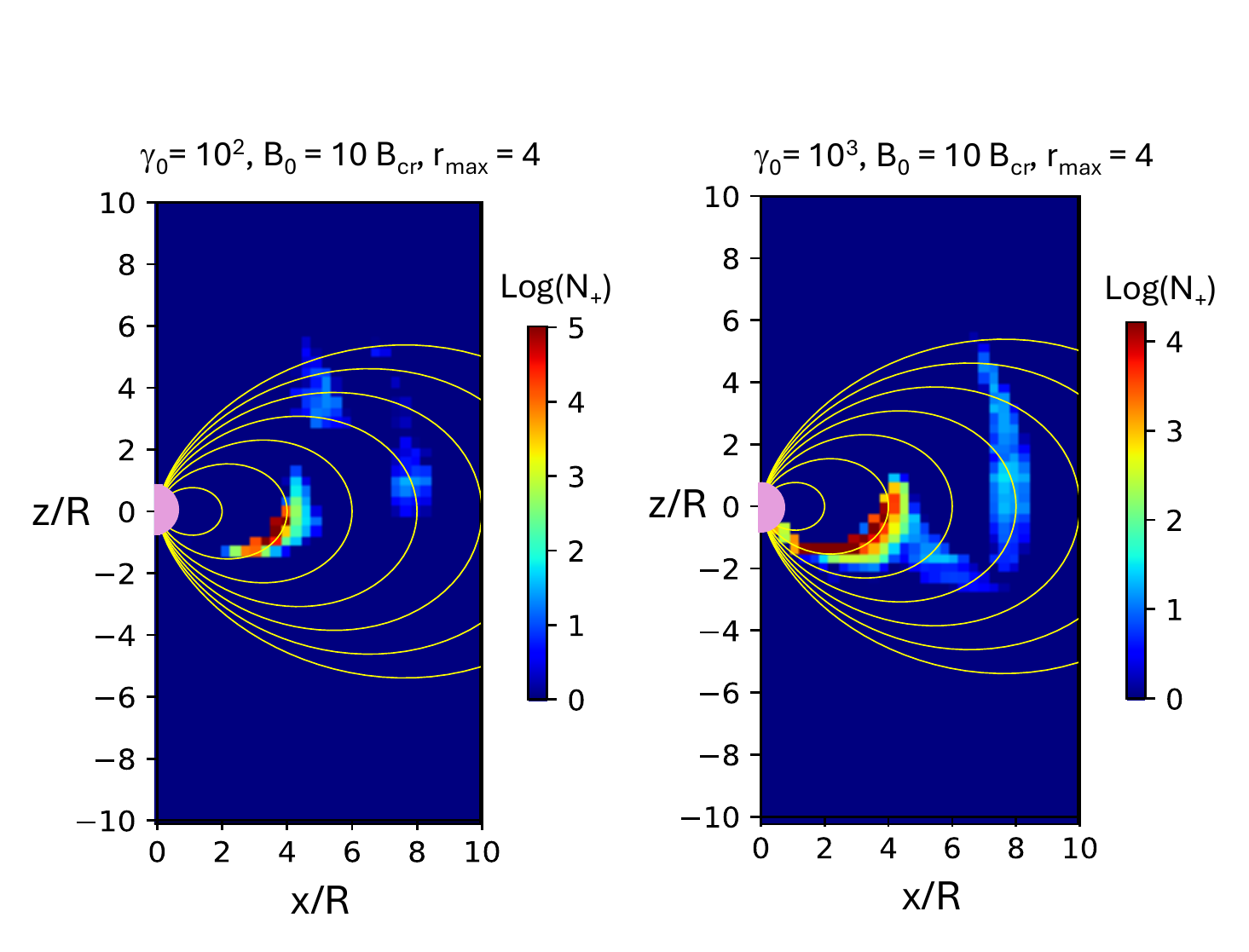}
\caption{Spatial distribution of multiplicity (color scale, base 10 logarithms) of electron-positron pairs at their production point in the meridional plane of a magnetic field of surface strength $B_0/B_{\rm cr} = 10$ with dipole axis along the z direction, from a pair cascade initiated by a primary electron with no RICS losses, having constant Lorentz factor $\gamma_0 = 10^2$ (left panel) and $\gamma_0 = 10^3$ (right panel) on a closed field loop with maximum radius, $r_{\rm max} = 4$.}
\label{fig:pairdistg3r4nl}
\end{figure}

Figure \ref{fig:pairdistg3r4nl} shows the spatial distribution of pair multiplicity in the dipole meridional (x-z) plane for the case $B_0/B_{\rm cr} = 10$, $r_{\rm max} = 4$ (i.e., footpoint colatitude $\theta_{\rm fp}=30^{\circ}$), $\gamma_0 = 10^2$ and $\gamma_0 = 10^3$ with no RICS losses.  In both cases, most pairs are produced close to the dipole field line on which the primary electron travels with higher pair generations spread to more outer field loops.  For $\gamma_0 = 10^2$ the pairs do not appear at radii less than about 2 NS radii but become more concentrated near the top of the loop. The reason for this delayed onset is because such a low $\gamma_0$ generates RICS emission only up to around $10-20\;$MeV (see Figure~5, left) so that pair creation mean-free paths are longer. For $\gamma_0 = 10^3$, the RICS emission extends above 100 MeV (see Figure~3, left) so that pairs appear very close to the NS surface since the photon mean-free paths are shorter.  Yet it is noticeable  that the pairs also become more concentrated near the top of the loop and spread further to outer field loops.  The pair distribution for the same $B_0$ and $\gamma_0$ values in the no RICS loss case are shown in Figure \ref{fig:pairdistg3r6nl} for a particle trajectory on a field loop with larger $r_{\rm max} = 6$, for which the overall pair multiplicity drops somewhat -- see Table~1.  In this case, the multiplicity distribution looks similar near the start of the trajectory but near the top of the loop is spread over a much larger number of field loops.  As the cascade spreads to larger radii, the local field drops, increasing the photon pair production mean-free paths and the pairs are produced in very high Landau states, generating  both a broader spectrum and a more extended spatial distribution of SR.  Although the pairs produced at larger radii have smaller energies (see Figure \ref{fig:pairspecNL} below), their RICS radiation, which we have not included, will still contribute to the total spectrum at low energies.  Finally Figure \ref{fig:pairdistg3r4} shows the pair multiplicity distribution for $\gamma_0 = 10^3$ and $r_{\rm max} = 4$ with RICS losses included, for two different surface field strengths, $B_0/B_{\rm cr} = 10$ and $B_0/B_{\rm cr} = 50$.  There are dramatically fewer pairs in these cases, since the radiation losses decrease the maxima of the unattenuated spectra (see Figure \ref{fig:RICSg3}). The pairs are also very concentrated along the particle trajectory along the active field line.  The total multiplicities are not that different for the two field strengths but the pairs are more concentrated near the outer part of the loop for $B_0/B_{\rm cr} = 50$ since the particles lose energy more slowly.  For both cases, the total multiplicities are small and there is a larger fraction at low Lorentz factor, so the neglect of RICS from pairs is a much better assumption.  

\begin{figure}[h] 
\includegraphics[width=190mm]{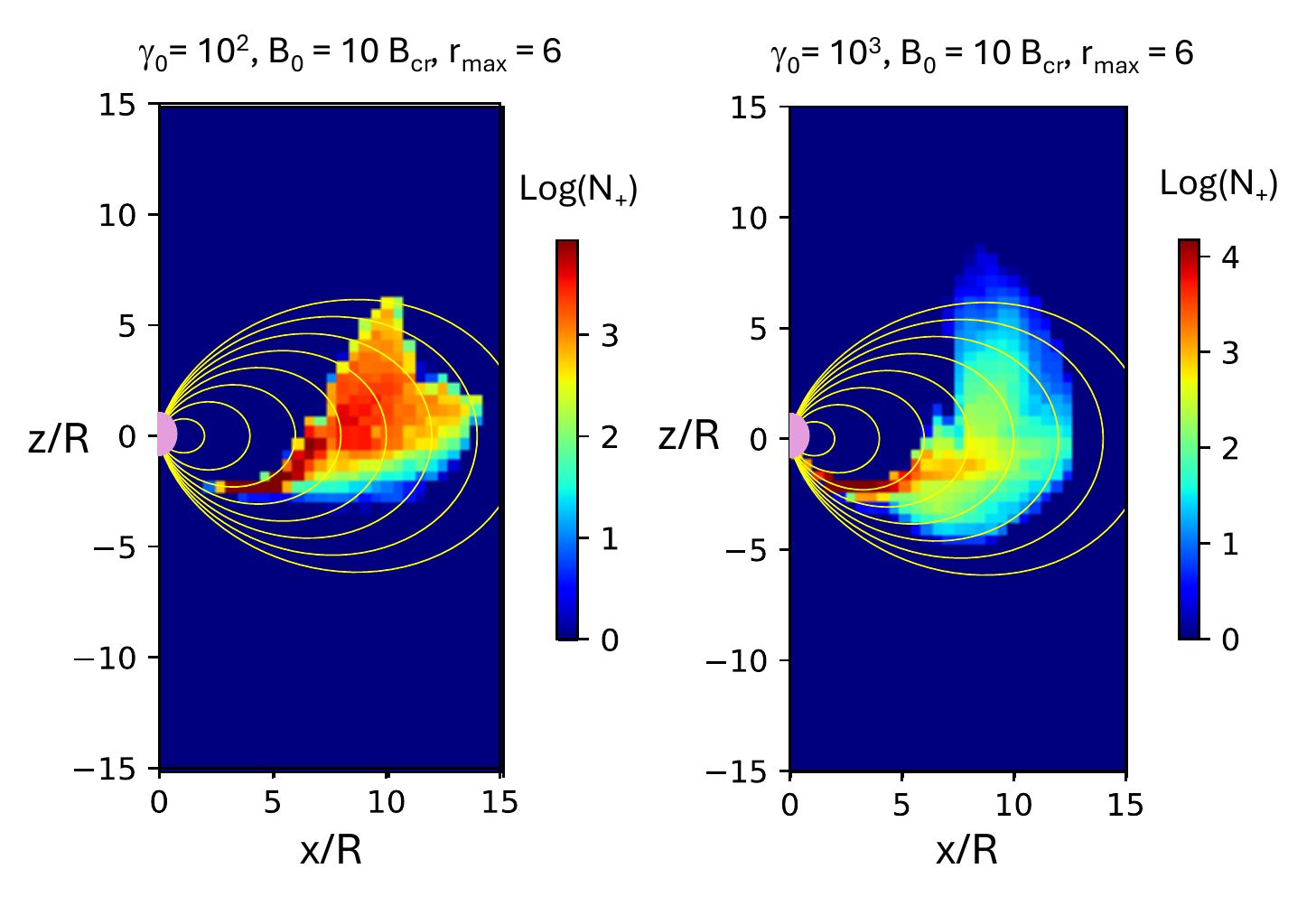}
\caption{Same as Figure \ref{fig:pairdistg3r4nl} for a closed field loop with maximum radius, $r_{\rm max} = 6$.}
\label{fig:pairdistg3r6nl}
\end{figure}

The spectral distribution of pair energies for the no RICS loss case is shown in the left panel of Figure \ref{fig:pairspecNL}.  The spectra extend almost to the primary particle Lorentz factor and look very similar for the two field strengths in the $\gamma_0 = 10^3$ case.  The $\gamma_0 = 10^2$ spectrum is much steeper with many more pairs at Lorentz factors less than 10.  The extra low energy pairs however give the highest multiplicity of all the cases in Table 1.  The pair spectra for the same parameters but for the RICS loss case is shown in the right panel of Figure \ref{fig:pairspecNL}.  With particle energy losses, the $\gamma_0 = 10^3$ spectrum for the $B_0/B_{\rm cr} = 10$ field strength is much higher than for the $B_0/B_{\rm cr} = 50$ case and has more than three times the multiplicity, since the RICS energy losses are lower for the high field case -- see Figure \ref{fig:dynamics}, right, noting that the scattering resonance is less accessible for such large surface fields.  The $\gamma_0 = 10^3$ spectrum is again steeper with the bulk of the pairs are the lowest energies, so that the multiplicity is comparable to the $\gamma_0 = 10^3$, $B_0/B_{\rm cr} = 50$ case.  With most of the pairs in the $B_0/B_{\rm cr} = 10$ spectra at Lorentz factors of a few, there will not be a significant pair RICS or cascade contribution to the total photon spectrum.  The cascade simulation of \citet{Belo2013b} for the case  $\gamma_0 = 10^3$, $B_0/B_{\rm cr} = 10$ and RICS energy loss gave an estimated pair multiplicity $\sim 100$, in good agreement with our value of 62 for this case.  However, our pair energy distribution, with an index of $\sim -1$, is steeper than what is shown in that paper, possibly because we allow pair production in excited states.

\begin{figure}[h] 
\includegraphics[width=180mm]{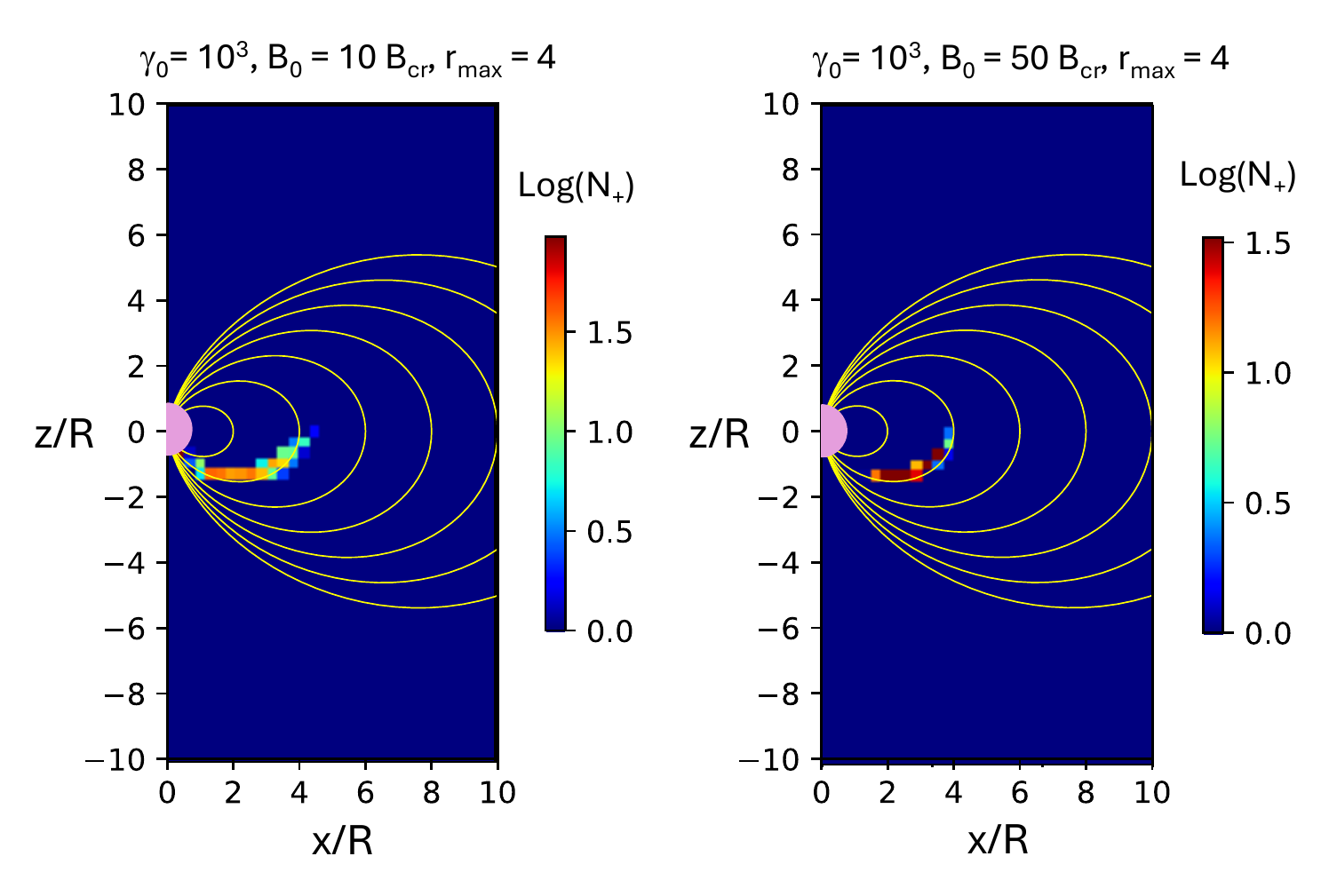}
\caption{Left panel: Same as Figure \ref{fig:pairdistg3r4nl} for a case where RICS energy losses are included in the primary electron dynamics.  Right panel: Same as left panel for surface magnetic field strength $B_0 = 50\,B_{\rm cr}$.}
\label{fig:pairdistg3r4}
\end{figure}

\begin{figure} 
\includegraphics[width=180mm]{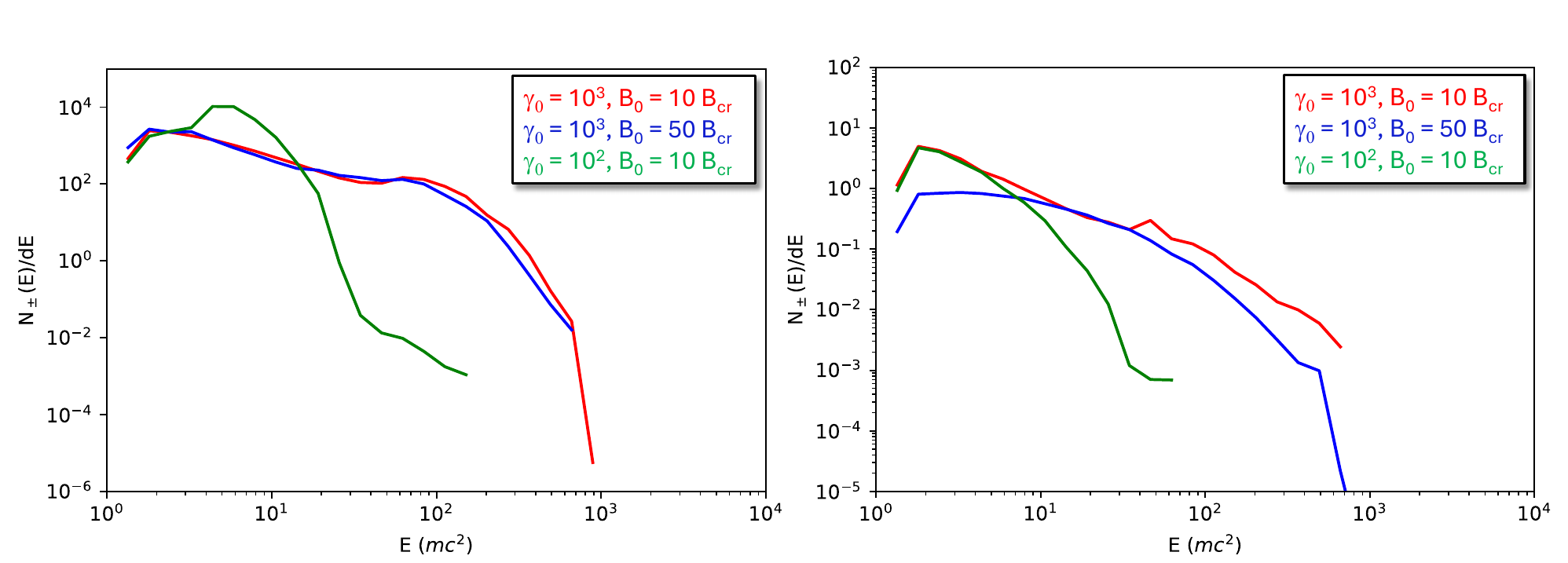}
\caption{Total electron-positron pair energy spectra from cascades on a closed dipole field loop of maximum radius, $r_{\rm max} = 4$ for different constant primary electron Lorentz factors and surface magnetic field strengths, as labeled.  Left panel: no RICS losses. Right panel: RICS losses.}
\label{fig:pairspecNL}
\end{figure}

\section{Discussion}  \label{sec:disc}

We have presented a Monte-Carlo simulation of pair cascades initiated by primary electrons injected at the base of closed dipole magnetic field loops of a magnetar magnetosphere.  Both cases where the primary electrons keep a constant Lorentz factor around the loop, and where the electrons continually lose energy to RICS emission, are considered.  In the first case, the acceleration would need to occur over the whole particle trajectory and just balance the RICS drag force.  This scenario is not likely since the pairs produced by the RICS photons will at least partially screen an accelerating electric field. This acceleration could be sustained on the local length scale associated with the skin depth of the plasma of primaries and secondaries. Additionally, the screening is spatially non-local owing to the strong angle dependence of pair creation, thus regions or layers with partially unscreened fields may survive. For low altitudes where the field is roughly constant, the competition between acceleration and RICS losses, as a function of Lorentz factor, can be seen in Figure~3 of \citet{Cooper2024}. In self-consistent models of pulsar polar cap pair cascades \citep{2010MNRAS.408.2092T,Timokhin2013}, the screening is non-steady with primary accelerated particles producing bursts of pairs that completely screen the electric field, followed by further cycles of acceleration, pair production and complete screening.  The end result is a non-steady beam of primary particles, whose energy is limited by pair screening, that produce a full cascade above the screening zone while suffering radiative losses \citep{Timokhin2015}.  In the second case, primary electrons are injected with high Lorentz factors at the base of the loop through a localized electric discharge \citep{Belo2007}, the scenario assumed in the Monte Carlo simulations of \citet{Belo2013b}.  In either case, the primary electrons end up losing energy to RICS drag over a large part of their trajectory.  The exception may be a case where the primary electron acceleration produces Lorentz factors below $10^2$, where the RICS loss rate is small or negligible through the whole trajectory \citep[see also][]{Baring2011,Cooper2024} and there are too few pairs to screen the electric field (see Table 1).  In the left-hand panel of Figure \ref{fig:dynamics}, the $\gamma_0 = 10^2$ case barely reaches the resonant condition near the top of the loop.  The RICS no-loss and loss cases result in very different pair multiplicity, for primary Lorentz factors $\gamma_0 \gtrsim 10^2$, with multiplicities $\gtrsim 10^4$ in the cases without RICS loss but $< 100$ in the RICS loss cases.  
 
In both cases, the RICS-induced cascade spectra are dominated by either split photons or SR from pairs.  For large viewing angles near the start of the electron trajectory, where the local magnetic fields are highest, split photon emission dominates over at least part of the spectrum.  At smaller viewing angles near the top of the loop, the pair SR dominates over most of the spectrum.  The split photon spectra have the same photon index as the RICS spectra, but the pair SR spectra are much softer since the pairs have a broad spectrum of energies both parallel to and transverse to the {\bf B} field. \citet{Belo2013b} assumed that all pairs are produced at threshold in the ground state, so that the spectra have no SR components.  However, we find that, while pairs are produced at threshold in the ground state near the start of the trajectory, pairs are produced in excited states as soon as the local field drops below $\sim 0.3\,B_{\rm cr}$.

In this initial investigation of magnetar pair cascades, we have made a number of simplifying assumptions.  The electron trajectory and all of the cascade radiation is assumed to lie in a single plane at one azimuth angle.  We also assumed that the RICS emission at all energies is radiated along the local magnetic field.  In reality, only the highest energies in the spectrum ($\varepsilon_f \gtrsim 0.1 mc^2$) are emitted at very small angles to the field (\citetalias{Wadiasingh2018}).  Therefore, observers at other azimuth angles would see no emission when they should be seeing the low-energy part of the spectrum.  However, according to Figure 7 of \citetalias{Wadiasingh2018}, there is no emission above a few $mc^2$ from a given loop at azimuthal angles $\gtrsim 5^\circ$, around where pair production and splitting cut off the spectra.  However, since we allow random azimuthal angles for the SR emission, which can be radiated by pairs with low Lorentz factors, the SR will be spread beyond the loop plane.  The emission at other azimuths will also depend on the azimuthal distribution of electron injection and if, as is likely, injection occurs over a range of azimuths, then the high-energy spectrum in each viewing direction will be dominated by the emission of a loop that is coplanar with that azimuth.  It may then be possible to use the results from a cascade on a single loop as presented herein to simulate spectra for a range of azimuths.  

We also have ignored the RICS emission from pairs, which we have shown should be small or negligible for cases with RICS energy losses.  For cases without RICS energy loss, the contribution of pair RICS would be substantial and dominate the primary cascade radiation.  However, we have argued that the case with constant Lorentz factor is not very likely, given that it requires a fine-tuned balance between RICS cooling and the acceleration associated with electrodynamic potentials/gaps in twisted magnetospheres.  Finally, we have not included effects of strong gravity near the NS, specifically redshifting and light bending of photons and GR effects on the dipole magnetic field \citep{1974PhRvD..10.3166P,Wasserman1983,1986SvA....30..567M,2013MNRAS.433..986P}.  These effects were previously studied in the context of photon pair production and splitting escape energies \citep{Harding1997,BaringHarding2001,Story2014,Hu2019}, the maximum photon energies permitting transparency.  The major influences are the $\sim 1.3$ photon redshift factor that effectively increases the photon energies at the NS surface compared with the observed energies, and a $\sim 1.4 $ increase in dipole magnetic field strength at the NS surface.  Both these effects decrease photon attenuation mean-free paths and lower the escape energies for pair production and splitting.  

We have also assumed that only the photon splitting mode, $\perp \rightarrow \parallel\parallel$, is operational.  Two other modes may be allowed in the regime of strong vacuum dispersion, which could apply to magnetar fields.  Wadiasingh et al. (2025) show RICS spectra attenuated by photon splitting if three modes are operating, and both photon polarization modes can split, circumventing pair production.  In the case of pair cascades for surface field, $B_0 = 10\,B_{\rm cr}$, splitting in three modes will prevent pair production at large viewing angles when the electron is in the highest fields but not near the top of the loop when pair production takes over.  But for $B_0 = 50\,B_{\rm cr}$, and $r_{\rm max} = 4$ , photon splitting dominates over the entire loop and there may be no pairs produced in the cascade.  Splitting in three modes would also soften the split photon spectrum and change the polarization characteristics near the cutoff.

This study also does not explore the influences of magnetic field twists on the photon attenuation.  Such toroidal modifications/additions to the basic dipolar field morphology need to be present in order to provide at least the initial acceleration of the particles \citep{Belo2007}.  Large twists, if they impart strong electric fields or small skin depths, giving a high acceleration rate in gaps or a very high initial Lorentz factor, may instead favor a curvature radiation channel over RICS for the initial pairs produced \citep{Baring2011,Cooper2024}.  Field twists both increase the magnetic field strength and straighten poloidal field lines compared to a pure dipole field configuration \citep{Thompson2002,Hu2022}.  In an extensive study of hard X-ray/soft gamma-ray opacities in twisted magnetospheres, \citet{Hu2022} observed that the impact of field twists on pair creation and photon splitting attenuation is intricately dependent on the competition between field enhancement (which increases the opacity) and field-line straightening (which decreases the opacity), and is determined by the location of photon emission. As noted in \citet{Hu2022}, the field line straightening implies a maximum permissible twist if the current carriers are supplied by pair production.  For the polar field strengths and fieldline footpoint colatitudes employed in the various cascade illustrations in this paper, Figures~4 and~5 of \citet{Hu2022} indicate that stronger twists increase the net opacity primarily due to the field enhancement, yet opacity for splitting tends to gain more than that for pair creation.  Accordingly, assessments of the net pair and accompanying SR yields for twisted field geometries will require dedicated simulation studies.

Although the RICS spectra in general are too hard to match the observed hard components of magnetars (without invoking trajectories spanning significant volumes, Wadiasingh et al. 2025), with photons indices of 1 - 1.5, the pair SR spectra have much softer photon indices that are in this range.  The SR components also have fluxes that are at least several orders of magnitude higher than the RICS spectral flux, essentially due to the energy re-apportionment in reprocessing soft gamma-ray RICS emission via pair creation.  For a source distance of $d\sim 3\;$kpc,  we estimate that primary particle injection levels of $10^{36} - 10^{37}\,\rm s^{-1}\,(d/3\,kpc)^2$ are required to match the observed spectral fluxes for the cascade SR spectra.  The pair SR spectra are also highly polarized with fractions of 40\% - 50\%, while the RICS spectra for primary Lorentz factors $\gtrsim 10^2$ are unpolarized except near the cutoff at the maximum energy.  Observations by the Imaging X-ray Polarimetry Explorer (IXPE), at energies from 2 - 10 keV, have shown that some magnetars have polarization increasing with energy.  \citet{Zane2023} detect a strong polarization signal from 1RXS J170849.0-400910 that reaches $57.7 \pm 6.8$\% at 5-6 keV and  $85 \pm 15$\% at energy 6 - 8 keV.  IXPE observations of another magnetar, 1E 1841–045 40 in outburst \citep{Stewart2025} also find an increasing polarization degree with energy with $51 \pm 10$\% for the hard component that dominates the spectrum at 4 - 8 keV.  Our SR spectra generally have polarization degree around 50\% but in some cases (Figures \ref{fig:caspecg3b10r4nl} and \ref{fig:caspecg3b50r4}) can reach 80\%.  

Alternative models have been proposed to explain the hard components of magnetars.  \citet{Thompson2020} propose annihilation Bremsstrahlung emission from dense, collisional pair plasma formed near the neutron star in regions of large current.  But this mechanism requires very high pair densities, $n_e \gtrsim 10^{20}$ cm$^{-3}$, for the emissivity to exceed RICS that are higher than the current needed for the magnetic twists.  Such densities yield emission regions that are highly optically thick to Compton scattering.  It is also not clear how such a dense, transrelativistic plasma can form and be maintained in a magnetar magnetosphere, and also avoid thermalizing and thereby narrowing the band of hard X-ray emission considerably. 

Our cascade studies are possibly relevant for fast radio burst models \citep[see][for a review]{2023RvMP...95c5005Z} and radio emission from magnetars \citep{2006Natur.442..892C,2017MNRAS.465..242T,2021MNRAS.502..127L}, particularly inner magnetospheric magnetar models that rely on pair creation for coherent radio emission \citep[e.g.][]{2019ApJ...879....4W,2020ApJ...891...82W,2023MNRAS.520.1872B,Cooper2024}. The pair luminosity, of the total cascade, sets an upper bound to the beaming-corrected luminosity of radio emission. We have shown that the pair cascades in magnetars are quite different than those for conventional pulsars.  The cascade pair multiplicity, energy distribution and locales where it occurs will influence generation and escape (including observed radio polarization) of coherent radio emission from the inner magnetosphere \citep[e.g.][]{2019ApJ...882L...9M,2024NatAs...8..606L} and the observed spectral index of the emission in a non-trivial manner. 
 
Looking ahead to the future, generating pulse profiles and phase-resolved spectra of the cascade emission will require modeling magnetospheres with inclined dipoles and assumptions about the distribution of injected primary electrons in azimuth and on different magnetic field loops.  Wadiasingh et al. (2025) have studied the integrated RICS emission from electrons injected uniformly at the base of collections of closed field loops and included the attenuation by photon splitting in three modes.  They found RICS pulse profiles generically narrow with higher observed photon energy, similar to that observed with INTEGRAL. They also found that the pure RICS spectra integrated over multiple field loops are softer than those from single field loops. The cascade spectra we have presented in this paper are dominated by split photons and pair SR.  Since the split photon spectra have the same spectral behavior as the RICS spectra, we expect such a softening to also be present in the split photon spectra when integrating over extended volumes.   Rotationally-modulated, volume-integrated cascade SR spectra are harder to predict but will depend mostly on the variation of the local magnetic field along the line-of-sight.  We plan to study cascades from collections of field loops and including photon splitting in three modes, exploring the pulse profiles and phase-resolved spectra from these configurations in a forthcoming paper.


\begin{acknowledgments}
The numerical computations for this work were partly performed on the Perlmutter cluster at the DOE National Energy Research Scientific Computing Center (NERSC).  
Z.W. acknowledges support by NASA under award numbers 80GSFC21M0002 and 80GSFC24M0006. 
M.G.B. acknowledges support from NASA under grant 80NSSC24K0589.
\end{acknowledgments}

\begin{appendix}

\begin{figure} 
\includegraphics[width=130mm]{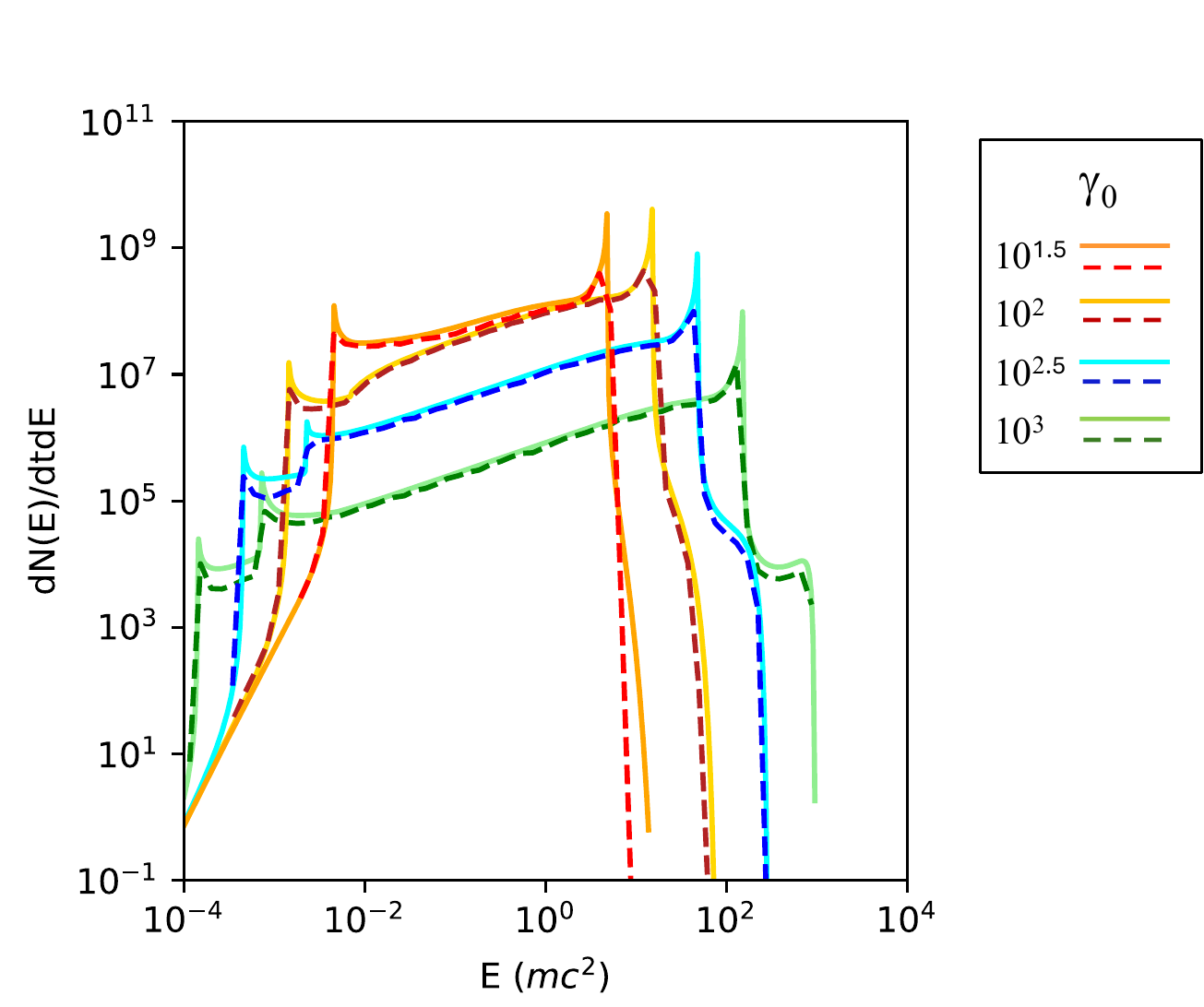}
\caption{Resonant inverse Compton scattering spectra from single electrons of constant Lorentz factor $\gamma_0$ on a single closed magnetic dipole field loop of maximum radius $r_{\rm max} = 4$ for observer angle  $\theta_v = 30^\circ$, surface magnetic field strength, $B_0/B_{\rm cr} = 10$, and NS surface temperature, $T = 5 \times 10^6$ K.  Solid lines are spectra from Figure 6 of \citet{Wadiasingh2018} and dashed lines are spectra computed from the code presented in this paper.}
\label{fig:compW2018}
\end{figure}

In order to verify the RICS spectral calculation part of our code, we have computed the spectra from a single electron with constant Lorentz factor (no RICS losses) on a single closed magnetic field loop for specific observer angles, $\theta_v$.  To compare with the results of \citetalias{Wadiasingh2018}, we normalize the spectra by the total distance over the whole loop. Figure \ref{fig:compW2018} shows a comparison of our spectra with those in Figure 6 of \citetalias{Wadiasingh2018}.  The agreement is very good over most of the spectrum except at the lowest energies, since our code uses the high Lorentz factor approximations of Equations (35) - (37) of \citetalias{Wadiasingh2018} to determine the two resonance interaction points on the loop that appear for the highest scattered photon energies.  \citetalias{Wadiasingh2018} compute all possible resonance points using bisection to determine the roots of their Equation (32), which captures the low-energy part of the spectrum more accurately where more than two resonance points appear.  Our code-calculated spectra do not fully capture the high-energy cusps as a result of our smaller number of photon-energy bins.  With the exception of the normalization, the spectral calculations for this plot are the same as used in the cascade simulation. 

\end{appendix}

\bibliography{Magcas}{}
\bibliographystyle{aasjournalv7}

\end{document}